\documentclass[a4paper,11pt]{article}
\pdfoutput=1 
\usepackage{jcappub}
\usepackage[utf8]{inputenc}
\usepackage{graphicx}
\usepackage{bm}
\usepackage{amsmath}
\usepackage{siunitx}
\usepackage{hhline}
\usepackage{multirow}
\usepackage{tabularx}
\usepackage{comment}
\usepackage[dvipsnames]{xcolor}
\usepackage{mathtools}
\usepackage[normalem]{ulem}

\newcommand{\planck}{\textit{Planck}}
\newcommand{\euclid}{\textit{Euclid}}

\begin{document}

\title{Testing gravity with gravitational wave friction and gravitational slip}

\author[a, b]{Isabela S. Matos,}
\author[c, d, e, b]{Emilio Bellini,}
\author[a]{Maurício O. Calvão}
\author[b]{and Martin Kunz}

\affiliation[a]{Universidade Federal do Rio de Janeiro, Instituto de F\'\i sica\\
CEP 21941-972 Rio de Janeiro, RJ, Brazil}
	
\affiliation[b]{Département de Physique Théorique and Center for Astroparticle Physics,
Université de Genève, Quai E. Ansermet 24, CH-1211 Genève 4, Switzerland}

\affiliation[c]{INFN, National Institute for Nuclear Physics, Via Valerio 2, I-34127 Trieste, Italy}

\affiliation[d]{IFPU, Institute for Fundamental Physics of the Universe, via Beirut 2, 34151 Trieste, Italy}

\affiliation[e]{SISSA, International School for Advanced Studies, Via Bonomea 265, 34136 Trieste, Italy}

\emailAdd{isa@if.ufrj.br}
\emailAdd{emilio.bellini@ts.infn.it}
\emailAdd{orca@if.ufrj.br}
\emailAdd{martin.kunz@unige.ch}

\abstract{{Gravitational waves (GWs) emitted by binary sources are interesting signals for testing gravity on cosmological scales since they allow measurements of the luminosity distance. When followed by electromagnetic counterparts, in particular, they enable a reconstruction of the GW-distance-redshift relation.} In the context of {several modified gravity (MG)} theories, even when requiring that the speed of propagation is equal to that of light, this GW distance differs from the standard electromagnetic luminosity distance due to the presence of a modified friction in the GW propagation. The very same source of this friction, which is the running of an effective Planck mass, also affects the scalar sector generating {gravitational} slip, i.e.~a difference between the scalar potentials, an observable that can be inferred from large-scale structure {(LSS)} probes. In this work, we use a model within effective field theories for dark energy to exemplify precisely the fact that, at the linear perturbation level, parametrizing a single function is already enough to generate simultaneous deviations in the GW distance and the slip. By simulating multimessenger GW events that might be detected by the Einstein Telescope in the future, we compare the constraining power of the two observables on this single degree of freedom. We then combine forecasts {of an \euclid-like survey} with GW simulations, coming to the conclusion that, when using Planck data to better constrain the cosmological parameters, those future data on the scalar and tensor sectors are competitive to probe such deviations from General Relativity{, with LSS giving stronger (but more model-dependent) results than GWs.}}

\maketitle

\section{Introduction}

Since the first detection of gravitational waves (GWs) by the LIGO-Virgo collaboration in 2015 \cite{Abbott2016a}, {a big effort has been devoted to {exploring} how much this new probe will help us to understand the Universe in various fields, such as astrophysics, cosmology and gravitation} \cite{Bailes2021}. {Particularly the} GW events that are followed by electromagnetic counterparts, such as gamma ray bursts, stand out as important for cosmology, since they provide measurements of both the redshift of the source and its luminosity distance, as first remarked in \cite{Schutz1986}. These so-called standard sirens constitute, therefore, a new class of sources that allows the reconstruction of the distance-redshift relation, complementary to type Ia supernovae surveys but without the need of tricky calibrations. In fact, the detection of multimessenger events has already yielded a measurement of the Hubble constant \cite{Abbott2021} which, even if not yet competitive in accuracy with the current best measurements, illustrates that these observations provide a powerful additional independent test of the concordance $\Lambda$-cold-dark matter ($\Lambda$CDM) model.

Joint events of GWs and electromagnetic (EM) signals are also powerful to measure deviations of General Relativity (GR), as demonstrated by the first detection of this kind, GW170817 \cite{LIGOGW170817, LIGO2017, LIGO2019}. That detection alone put a stringent bound on the speed of propagation of GWs at low redshifts, allowing for a relative difference to the speed of light of just one part in $10^{15}$. {Apart from UV cutoff arguments \cite{deRham2018},} this information highly disfavored, if one wants to avoid fine-tuning issues, a {significant fraction} of the gravity theories within the Horndeski family \cite{Horndeski:1974wa, Deffayet2011, Kobayashi:2011nu}, one of the most general Lagrangians that give rise to second order field equations with a single scalar field.

However, there is still a wide range of models not yet ruled out by observations, within and beyond Horndeski, that feature modifications in the propagation of GWs through cosmological distances even with a speed equal to the speed of light \cite{Baker2017, Creminelli2017, Ezquiaga2017, Sakstein2017}. In particular, they change the friction rate at which the tensor modes propagate, an effect that can be encoded as a modification on what has been called recently the GW distance \cite{Lombriser2016, Amendola2018, Belgacem2018G, Nishizawa2018}. Since the GW signals exhibit a radiative behavior far from the sources, i.e. their amplitude is inversely proportional to the distance, this GW distance can be inferred from the detected waveform at ground or space-based interferometers, especially when the inclination of the binary source orbital plane is also relatively well determined. Such information might be  more precisely obtained in the future with the advent of a network of third generation detectors \cite{Maggiore2020, Amaro-Seoane2017, Reitze2019, Zhao2018, Arai2018}, covering a large range of frequencies and types of sources, that will also be sensitive to other signatures of modified gravity (MG), such as the existence of extra polarization modes \cite{Takeda2019}.

In this work we will focus on the Einstein Telescope (ET), a proposed underground cryogenic third generation GW observatory \cite{Maggiore2020, Sathyaprakash2010}. ET is particularly interesting for testing GR because, among other kinds of signals from sources up to redshift of order 20, those emitted by binary neutron stars (BNS) {fall into} its sensitivity frequency band ($\sim 1 - 10^4$ Hz). ET is expected to see, within a decade of running, several hundred multimessenger events emitted by BNS, up to redshift $z = 2$. Several works such as \cite{Matos2021,DAgostino2019,Belgacem2019b,Zhao2011,Nunes2019,Bachega2020, Nishizawa2019, Zhang2019, Jin2022} have recently investigated the ability of  ET to constrain gravity and cosmology, finding, for instance, that bright standard sirens will give a percent level measurement of the Hubble constant. Concerning MG, its effect on the GW distance is however highly degenerate with the cosmological background and for some particular models, even in the best scenarios, the results can be much weaker than current constraints (for example in $f(R)$, GWs will give $|f_{R0}| < 10^{-2}$ at best \cite{Matos2021}).

The degeneracy between MG and cosmology in the GW measurements requires combining them in the future with other powerful tests, such as observations of the large-scale structure of the Universe \cite{Mukherjee2021,Baker2021, Jin2021, Libanore2021}. A common way of currently testing gravity in the scalar sector is by measuring the gravitational slip, i.e. a difference between the two gravitational potentials $\Phi$ and $\Psi$, that is in general non-vanishing whenever we are in MG (see for instance \cite{Amendola2013,Amendola2014, Pinho2018, Blanco2021}). Indeed, measuring the slip by looking at the clustering of matter from the distribution of galaxies and the cosmic shear generated by weak lensing is one of the targets of galaxy surveys such as the upcoming \euclid~satellite \cite{Amendola2018b}.

The fundamental fact that will be explored in this work is that, in many MG theories such as the Horndeski models, the very same function that modifies the GW friction, which is the running of the Planck mass, also affects the scalar perturbations, generating slip. When aiming for model-independent, {phenomenological} tests of gravity, this relation is usually tuned out and the two observables are considered independently. The goal of this work is, on the contrary, to take advantage of {this} connection and to compare the constraining power of the scalar and tensor sectors of the cosmological perturbations on MG through measurements of, respectively, the slip and the GW distance, as first proposed in \cite{Saltas2014}.

The structure of the paper is the following. In Section \ref{sec:theory} we briefly introduce our theoretical framework of a linearly perturbed FLRW metric in MG theories, define our main observables (the GW distance and the slip) and the gravity models we studied. In Section \ref{sec:data}, we describe the analysed data sets: forecasts for a \euclid-like survey with both weak lensing and galaxy clustering contributions, cosmic microwave background (CMB) data from \textit{Planck} and simulations of GWs emitted by BNS detected by the Einstein Telescope. Finally, in Section \ref{sec:results} we present our results: first, a naive comparison between the constraining power of data from the tensor and scalar sectors on the Planck mass running rate; then the constraints on the cosmological parameters and our MG model from all data sets and their combinations. In Section \ref{sec:conclusion} we discuss these results and conclude.

\section{Theory}
\label{sec:theory}

Our cosmology is defined by a perturbed spatially flat FLRW metric,
\begin{equation}
ds^2 = a(\tau)^2\left[ -(1 + 2\Psi)d\tau^2 + (1 - 2\Phi)(\delta_{ij} + h_{ij})dx^idx^j\right]\,,
\end{equation}
where $a$ is the scale factor, $\tau$ is the conformal time, $\Phi$ and $\Psi$ are the scalar potentials and $h_{ij}$ is the transverse traceless tensor perturbation, that can be decomposed into $h_+$ and $h_{\times}$ polarizations.

In the context of cosmology, it has been shown \cite{Gubitosi:2012hu,Bellini2014,Gleyzes:2014qga} that the information content of a broad class of DE/MG theories  can be described at background and linear perturbation levels by only five functions of time,
\begin{equation}
\left({\mathcal H}, \alpha_M, \alpha_T, \alpha_K, \alpha_B\right)\,,
\end{equation}
where ${\mathcal H}:=a'/a$ is the conformal Hubble parameter $(a' := da/d\tau)$ and the $\alpha_i$'s are property functions that govern the evolution of both scalar and tensor perturbations. This effective field theory setting is equivalent, on cosmological scales, to one of the most general theories that give rise to second order field equations of motion with a single scalar field $\phi$, the Horndeski theory \cite{Horndeski:1974wa, Deffayet2011, Kobayashi:2011nu}, which includes families of models such as $f(R)$ \cite{Carrol2004, DeFelice2010} and quintessence \cite{Ratra1988, Wetterich1988}. The property functions $\alpha_i$ can be either inferred from a first-principled approach based on a particular Lagrangian or modeled with a suitable phenomenological parametrization;\footnote{See \cite{Bellini2014} for an in-depth discussion of their properties.} in both cases they encode possible deviations from the standard GR concordance cosmological model.

As first pointed out in \cite{Saltas2014}, two of these functions affect simultaneously the difference between the scalar potentials and the propagation of GWs through cosmological distances: the Planck mass run rate
\begin{equation}
\alpha_M := \frac{d(\ln M_{\ast}^2)}{d\ln a}
\end{equation}
and the tensor speed excess $\alpha_T$. More precisely, the GW polarization modes evolve in such theories according to
\begin{equation}
h_P'' + (2 + \alpha_M){\mathcal H}h'_P + (1 + \alpha_T)k^2h_P = \Pi_P\,, \label{GW}
\end{equation}
with $P = +, \times$, and where $\Pi_P$ is the tensor part of the anisotropic stress of matter, which is affected only by radiation and neutrinos and can be neglected at late times. The propagation equations show that, besides a non-luminal speed of propagation whenever $\alpha_T \neq 0$, the friction rate at which the modes are damped is modified by $\alpha_M$, changing the amplitude of the GW detected by local interferometers \cite{Nishizawa2018}. This gives rise to a new cosmological distance that is inferred from the detected GW waveform, the gravitational wave distance $D^{\text{gw}}$\cite{Belgacem2018, Amendola2018}. Its ratio to the standard electromagnetic luminosity distance in a given theory is given by
\begin{equation}
\Xi(z) := \frac{D^{\text{gw}}(z)}{D^{\text{em}}_L(z)} = \exp\left\{\frac{1}{2}\int_0^z \frac{\alpha_M(\tilde{z})}{1 + \tilde{z}}d\tilde{z}\right\} = \sqrt{\frac{M_{\ast}^2(0)}{M_{\ast}^2(z)}}\,. \label{gw_distance}
\end{equation}

For the scalar sector, on the other hand, among the equations that govern the evolution of the metric and matter perturbations, it is interesting to look at the one that gives the difference between the scalar potentials, which vanishes in GR when the scalar anisotropic stress of matter, $\Pi$, is negligible, which is normally the case at late times and inside the sound horizon,
\begin{equation}
   \Psi - (1 + \alpha_T)\Phi + (\alpha_M - \alpha_T){\mathcal H}\frac{\delta \phi}{{\phi}'} = \Pi\,, \label{phi_minus_psi}
\end{equation}
where $\delta \phi$ is the perturbation of the scalar field in the Newtonian gauge. Remarkably, as shown in \cite{Amendola2014}, it is possible to obtain in an operational way from observations, the slip parameter, defined as
\begin{equation}
    \eta := \frac{\Phi}{\Psi}\,,
\end{equation}
constituting an important variable in tests of MG \cite{Motta2013}.

The crucial fact that we will explore in this work is that in a given theory of modified gravity or dark energy, the deviation of the GW propagation and the gravitational slip from the GR values are generally linked \cite{Saltas2014, Nishizawa2019}, since both are described by equations that come from the spatial part of the Einstein field equations. We can see this explicitly in the example of effective field theories that we study here, where the same property functions affect GW propagation and gravitational slip.
Therefore, this motivates investigating whether measurements of the GW distance and the slip might be complementary (or not) to the inference of $\alpha_M$ and $\alpha_T$.

On the GW side, the link between $\alpha_M$ and the observable is given by Eq.\ \eqref{gw_distance} and is relatively simple. The connection between the gravitational slip and the $\alpha_i$'s is more complex as it depends on the evolution of the scalar field. In the quasi-static limit for $\alpha_T=0$ and assuming a $\Lambda$CDM background, following the notation of \cite{Bellini2014}, the slip is given by
\begin{equation}
    \eta = 1 - \frac{2\alpha_M(\alpha_B + 2\alpha_M)}{4(\alpha_B + 2\alpha_M) -2\alpha_B(1 + w_m)\tilde{\rho}_{\textrm{m}}/H^2 + 4\alpha_B'/\mathcal{H}} \, . \label{slip}
\end{equation}
We notice that for $\alpha_M = 0$ the slip vanishes, but in general it is a function of not only $\alpha_M$ but also $\alpha_B$. In the quasi-static limit the slip also vanishes for $\alpha_B = -2\alpha_M$, a case that is called the `no-slip' scenario \cite{Linder2018}.

In general it is not clear which parameter, $\alpha_M$ or $\alpha_B$, is more strongly constrained by a measurement of the slip. Only if the braiding parameter varies slowly it is possible to conclude that, in a DE dominated era, $\eta \approx 1-\alpha_M/2$ and that the slip is mostly constraining $\alpha_M$. This also illustrates that in general we would need to measure both the slip and the GW distance to disentangle the two parameters.

Another useful way to phenomenologically describe deviations from GR for linear scalar perturbations consists in modifying the Poisson and lensing potential equations \cite{Kunz:2012aw},
 \begin{align}
    k^2\Psi &= - \mu(a, k) 4\pi Ga^2\rho \Delta\,, \\
    k^2(\Phi + \Psi) &= - \Sigma(a, k) 8\pi Ga^2 \rho \Delta\,, \label{lensing}
\end{align}
where $\Delta$ is the gauge invariant density contrast as defined in \cite{Zucca2019}. The functions $\mu$ and $\Sigma$ are sometimes called $G_\textrm{matter}$ and  $G_\textrm{light}$ \cite{Linder2018} and probe modifications in, respectively, the growth of structure and the lensing potential. Although it is straightforward to obtain the slip from these functions, in the approach where they are independently parametrized, one loses the connection with GW propagation. This is why here we chose to parametrize the property functions instead, allowing us to track MG effects on the slip and the GW distance in a unified way.

\subsection{Base gravity model} \label{subsec:base_model}

We now define the base gravity model of our main analysis. First, as discussed before, in order to have a GW speed equal to unity we will set $\alpha_T = 0$. Second, we will use the relation $\alpha_B = - \alpha_M$, which corresponds to imposing a conformal coupling and that can be equivalently stated in terms of the Horndeski Lagrangian or the EFT base functions, as detailed in Appendix \ref{app:horndeski}.

The last property function $\alpha_K$ is then chosen such that the speed of sound of the scalar field, in the quasi-static limit, equals 1. This last choice should not impact our results strongly since, firstly, both the GWs and the slip in the quasi-static limit do not depend on $\alpha_K$, and secondly Refs.\ \cite{Bellini:2015xja,Gleyzes:2015rua,Alonso:2016suf} claim that $\alpha_K$ is hardly measurable. We end up with a model which is completely specified by a single free property function $\alpha_M$ (or, equivalently, the Planck mass) and the background evolution $H(\tau)$, which will be set to that of $\Lambda$CDM, since we are interested in the perturbations. We then parametrize the evolution of the Planck mass as a power law of the scale factor, so that the modifications of gravity kick in at late times, i.e.
\begin{equation}
    M_{\ast}^2 = 1 + \Omega_0 a^{\beta_0}\, , \label{parametrization}
\end{equation}
where $\Omega_0$ is a free amplitude parameter. This model, thus, recovers the physics of GR in the early Universe, where $a \rightarrow 0$.

\begin{figure}
    \centering
    \includegraphics[scale=0.49]{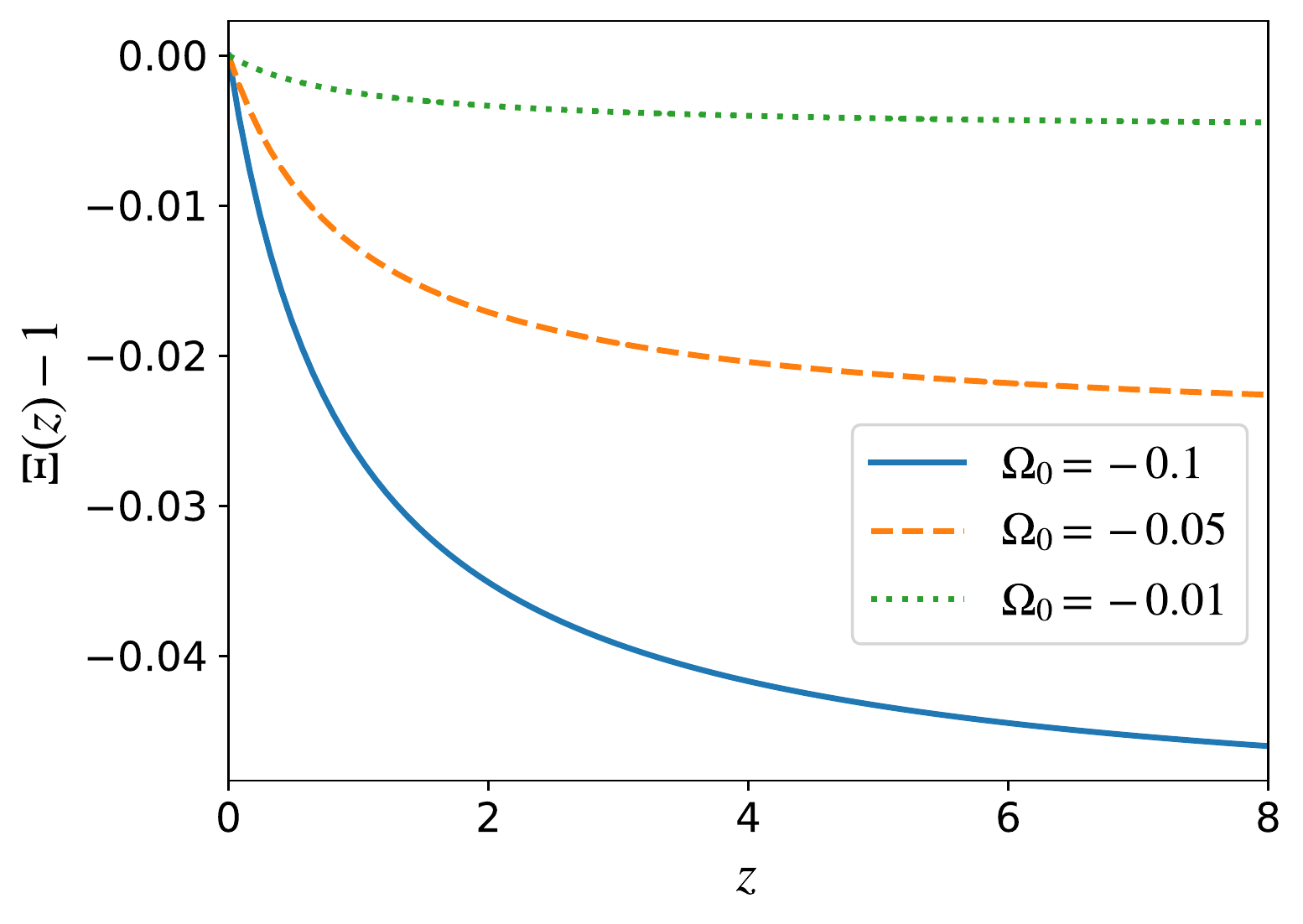}
    \includegraphics[scale=0.49]{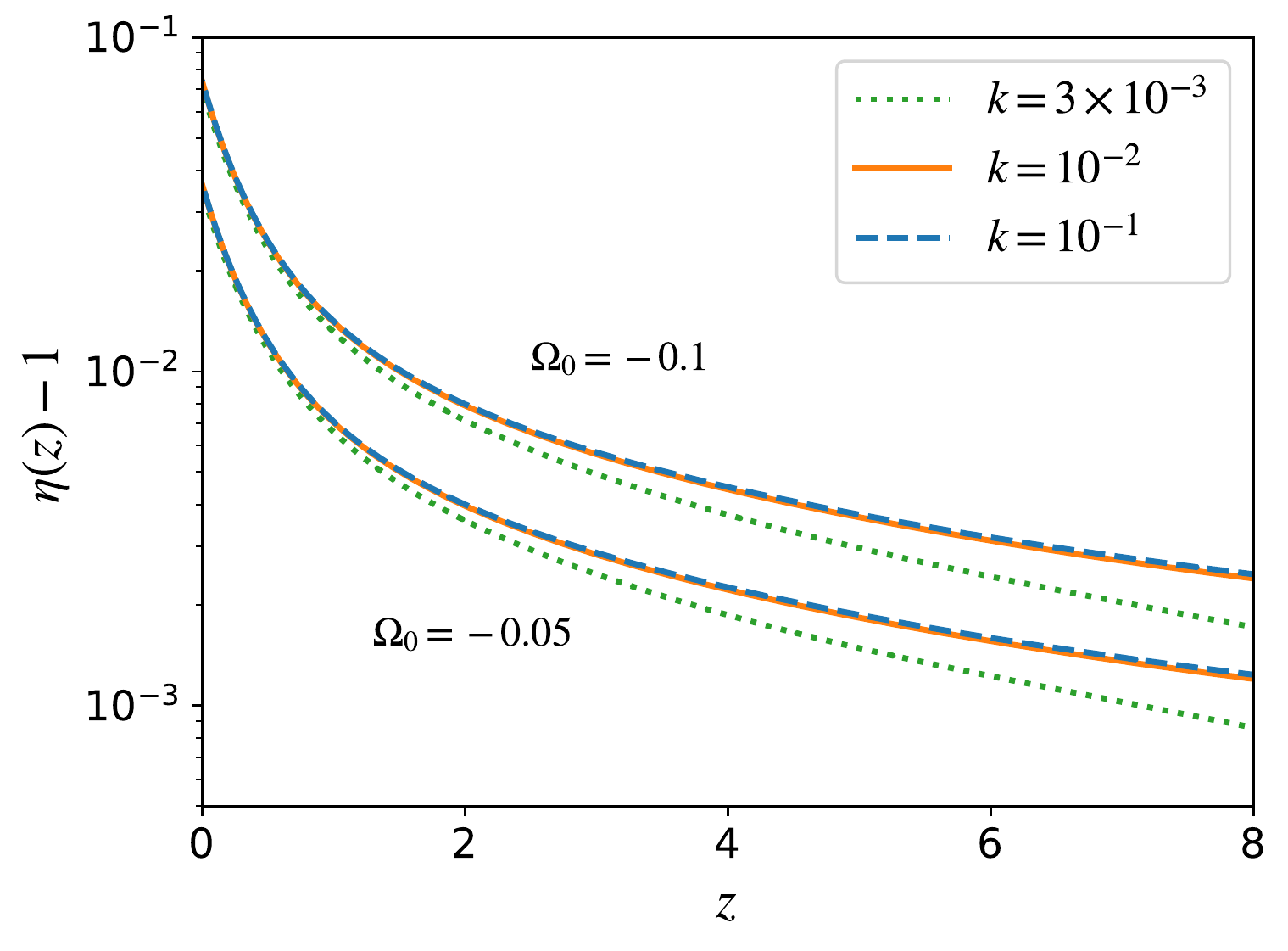}
    \caption{Left panel: The ratio between the GW and the luminosity distances that follows from the parametrization of the EFT base functions. The blue curve corresponds to our fiducial model, reaching a maximum 5\% deviation from the GR value of unity at the highest redshifts. On the right, the slip as a function of redshift for our fiducial model at different scales.}
        \label{fig:xi}
\end{figure}

We now briefly comment on the impact of such modifications of gravity on cosmological observables. Concerning the tensor sector, the above parametrization implies the ratio between the GW and electromagnetic luminosity distances in Eq.\ (\ref{gw_distance}) to be given by
\begin{equation}
\Xi(z) = \sqrt{\frac{1 + \Omega_0}{1 + \Omega_0 a^{\beta_0}}}\,. \label{xi}
\end{equation}
When the deviation from GR is small, i.e. $|\Omega_0| \ll 1$, it reduces to the parametrization proposed in \cite{Belgacem2018},
\begin{equation}
\Xi(z)  = \Xi_0 + \frac{1 - \Xi_0}{(1 + z)^n}\,,
\end{equation}
with $n = \beta_0$ and $\Xi_0 = 1 + \Omega_0/2$. The functional form in Eq.\ (\ref{xi}), which is depicted in Figure \ref{fig:xi}, mimics the typical behavior of $\Xi$ in $f(R)$ but also in several other MG theories. Its value at high redshifts depends only on $\Omega_0$ and so we fix $\beta_0 = 1$. Therefore, the models considered here feature only one additional degree of freedom $\Omega_0$, changing simultaneously the scalar and tensor sectors via the slip and the GW distance.

Regarding the scalar perturbations, such models differ from $\Lambda$CDM in the low-multipole temperature power spectrum $C_{\ell}^{TT}$ of the CMB and in its lensing potential power spectrum, which is roughly amplified by a global factor, as we will discuss in section \ref{sec:results} (cf. Figure \ref{fig:lensing_cls}). Also the matter power spectrum today in the linear regime, for scales inside the sound horizon, is simply multiplied by a constant factor, an effect which is degenerate with the galaxy bias (and also $\sigma_8$ or $A_s$). Finally, the slip parameter as a function of redshift is shown in the right panel of Figure \ref{fig:xi}. For our fiducial model, its value today is increased by about 7\%. We remark that for the GWs the maximum effect of the modification of gravity occurs at the highest redshifts, until it saturates, while the slip peaks today. However, this behavior of $\eta$, as compared to $\Xi$, is more sensitive to the exact parametrization assumed for $\alpha_M$, since they have roughly a proportionality relation, while the GW distance is an integrated effect. 

Finally, for the initial conditions at high redshift, we consider purely adiabatic scalar primordial perturbations with a power-law spectrum. We fix the primordial spectral index $n_s$, the baryon energy density $\omega_b$ and the reionization optical depth $\tau$ to the values shown in Table \ref{tab:fiducial_model}, corresponding to the best fit of the \textit{Planck} primary spectra from \cite{Planck2018}. We assume three neutrino species with only one massive state with mass $m_{\nu} = 0.06$ eV. The free parameters characterizing our models are, then,
\begin{equation}
    (\omega_c, A_s, h, \Omega_0)\,,
\end{equation}
which stand for, respectively, the cold dark matter density, the amplitude of the initial curvature perturbations at $k = 0.05 \; \mathrm{Mpc}^{-1}$, the dimensionless Hubble constant and the proportionality constant related to the Planck mass.

\subsection{No slip gravity} \label{subsec:no_slip}

Additionally, we investigate the no-slip model proposed in \cite{Linder2018}. These models have $\alpha_B = -2\alpha_M \neq 0$ but still no modification of the GW speed ($\alpha_T = 1$). They are interesting theories to compare our main analysis with because here only the propagation of GWs is modified in the quasi-static limit, since inside the horizon at late times
\begin{equation}
\mu = \Sigma = \frac{1}{M_{\ast}^2}\,,
\end{equation}
resulting in $\eta = 1$, as can be seen from Eq.\ \eqref{slip}.

\begin{figure}
    \centering
    \includegraphics[scale=0.49]{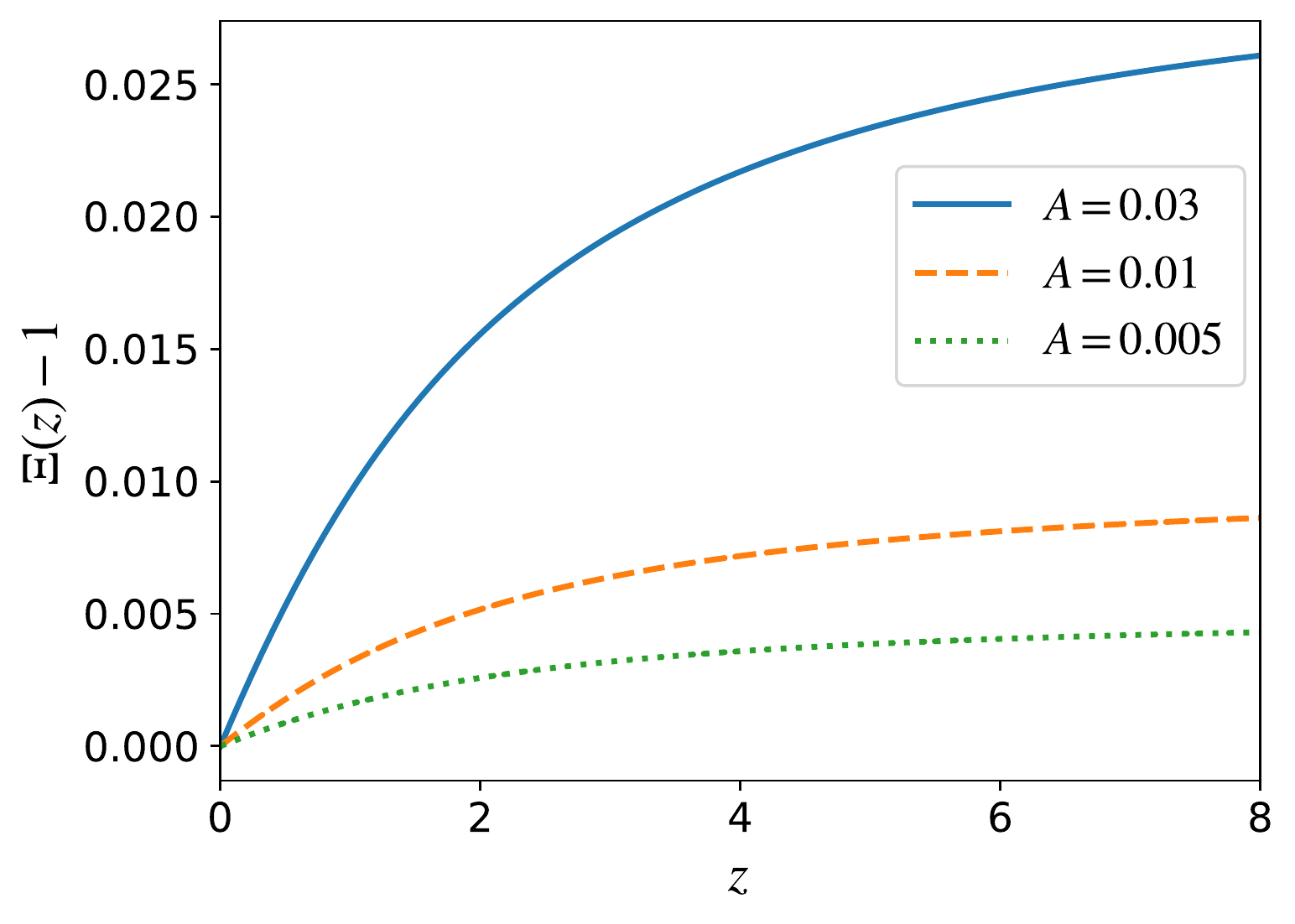}
    \caption{The ratio between the GW and luminosity distances as a function of redshift for the no-slip model with $a_t = 0.5$ and $r = 3/2$. The model with $A = 0.005$ is the fiducial model in our simulations while the one with $A = 0.03$ is discussed in \cite{Linder2018} and plotted here for comparison.}
    \label{fig:xi_no_slip}
\end{figure}

In this model, we adopt the following parametrization for the Planck mass and the corresponding running rate:
\begin{align}
    M_{\ast}^2 &= \textrm{e}^{(2A/r)(1 + \tanh{[(r/2)\ln(a/a_t)]})}\,,\\
    \alpha_M &= \frac{4A(a/a_t)^r}{[(a/a_t)^r + 1]^2}\,.
\end{align}
In terms of $\alpha_M$, this parametrization implies a `hill-like' behavior for the GW friction $\alpha_M$, as the viability conditions for this model ($c_s^2 \geq 0$ and no ghosts) demand that $\alpha_M \rightarrow 0$ both in the asymptotic past and future, while the modification reaches a peak value of $A$ at $a=a_t$. On the Planck mass side, $M_{\ast}^2$ starts with the value of unity at high redshifts up to the transition point $a = a_t$ after which it goes asymptotically to $\textrm{e}^{4A/r}$, with rapidity $r$. As compared to our base model, this behavior of the Planck mass in no-slip gravity gives an opposite deviation for the ratio between the GW and luminosity distances, i.e. it increases with redshift, as illustrated in Figure \ref{fig:xi_no_slip}

As discussed in \cite{Linder2018}, such a model with $A = 0.03$, $r = 1.5$ and $a_t = 0.5$ is compatible with measurements of the growth of structure observable $f\sigma_8$ when assuming a $\Lambda$CDM background. Therefore, we will systematically fix $r$ and $a_t$ to these values and explore the amplitude of the deviation from GR, which is encoded in the parameter $A$. By also setting $c_s^2 = 1$ we effectively have once again a one-parameter family of MG models.

\begin{table}[t] 
	\centering

	\setlength{\extrarowheight}{5pt}
	
	\begin{tabularx}{0.5\textwidth}{>{\centering\arraybackslash}X >{\centering\arraybackslash}X | >{\centering\arraybackslash}X >{\centering\arraybackslash}X}
		\hhline{= = = =}
		\multicolumn{2}{c}{Free}   & \multicolumn{2}{c}{Fixed}  \\ \hline
		$\omega_{c}$ & 0.1202 & $\omega_b$ & 0.02236 \\ 
		$10^9 A_s$ & 2.101 & $\tau_{\text{reio}}$ & 0.0544 \\ 
		$h$ & 0.6727 & $n_s$ & 0.963 \\ 
		$\Omega_0$ & $-0.1$ & $\sum m_{\nu}$ & 0.06 eV \\[2mm]
		\hline
		\end{tabularx}
	\caption{Parameter values of the fiducial model corresponding to those in Table 2 of \cite{Planck2018} (TT,TE,EE+lowE column), with the addition of the modified gravity related parameter $\Omega_0$.}
		\label{tab:fiducial_model}
\end{table}

\section{Data}
\label{sec:data}

\subsection{Gravitational waves}

We now briefly describe the procedure to simulate the GW mock data for the Einstein Telescope, as was done in \cite{Matos2021}.
We consider the waveform emitted at redshift $z$ by a BNS inspiral up to the 3rd post-Newtonian (PN) correction \cite{Blanchet2014}. It is given by:
\begin{eqnarray}
& \tilde{h}(f) = \mathcal{A} \sqrt{F_+^2 (1 + \cos^2\iota)^2 + 4F_{\times}^2\cos^2\iota} (f^{-7/6})e^{i\Phi(f)}\,, \\
& \mathcal{A} \coloneqq \sqrt{\dfrac{5}{96}}\dfrac{M_{c}^{5/6}}{\pi^{2/3}D^{\textrm{gw}}}\sum\limits_{j = 0}^{6}A_j \left(\pi Mf\right)^{\frac{j}{3}}\,,
\end{eqnarray}
where $M_c := (1 + z)M\left( \frac{\sqrt{m_1m_2}}{M}\right)^{3/5}$ is the redshifted chirp mass with $M = m_1 + m_2$ and
\begin{eqnarray}
F_+ \coloneqq  \frac{\sqrt{3}}{2}\bigg[\frac{1}{2}(1 + \cos^2\theta) \cos(2\phi)\cos(2\psi) - \cos\theta \sin(2\phi)\sin(2\psi)\bigg]\,, & \nonumber \\
F_{\times} \coloneqq  \frac{\sqrt{3}}{2}\bigg[\frac{1}{2}(1 + \cos^2\theta) \cos(2\phi) \sin(2\psi) + \cos\theta \sin(2\phi)\cos(2\psi)\bigg]\,, &
\end{eqnarray}
are the antenna pattern functions \cite{Zhao2011} when the angle between the arms {of the interferometer} equals $\pi/3$. Here, the parameters of the source $\bm{s} = (m_1, m_2, \iota, \theta, \phi, \psi)$ appearing in the above expressions stand for, respectively: the masses of the binary components, the angle of orbital inclination, the direction of the line of sight and the GW polarization angle. We assume, for simplicity, the spins of each one of the binary components to vanish in the coefficients $A_j$ of the PN expansion, whose expressions can be found in \cite{DAgostino2019}. The specific form of the phase function $\Phi$ does not affect our further calculations (cf. Eq.\ \eqref{SNR}).

The inference of the parameters encoded in the waveform, up to some degeneracies, can be made by a matched filtering procedure, where one compares several templates for $\tilde{h}$ with the observed signal considering a given noise spectrum \cite{Maggiore2007}. For high signal to noise ratio (SNR) events, one can work in the Fisher matrix approximation, where the instrumental error in the GW distance ends up depending only on the SNR evaluated at the most likely $\tilde{h}$, given by
\begin{equation}
\text{SNR}^2 = 4 \int_{f_{\text{low}}}^{f_{\text{up}}} \frac{\left|\tilde{h}(f)\right|^2}{S_n(f)}df\,. \label{SNR}
\end{equation}
Here $S_n$ is the noise power spectral density of the detector, which for ET we assume the ET-B model \cite{Zhao2011}, and $(f_{\text{low}}, f_{\text{up}}) = (1\textrm{ Hz}, 10^4\textrm{ Hz})$ are the limits of {ET's sensitivity band.  However, the distance and the orbital angle of the source are degenerate, effect that can be roughly taken into account by adding a factor 2 to the error \cite{DAgostino2019}, which becomes}
\begin{equation}
\sigma_{\text{ins}} = \frac{2 D^{\textrm{gw}}}{\text{SNR}}\,. \label{sigma_ins}
\end{equation}
We assume that we can use the geometrical optics limit for the GWs of interest to us, so that they propagate on null geodesics and are affected by gravitational lensing. This contributes an additional scatter to the GW distance that needs to be added in quadrature to the instrumental error, similarly to the treatment of supernova distance measurements. We use \cite{Sathyaprakash2009}
\begin{equation}
\sigma_{\text{lens}} = 0.05 z D^{\textrm{gw}}\,, \label{sigma_lens}
\end{equation}
which is a good approximation for the rigorous result of \cite{Martinelli2022} up to $z = 2$, giving a total error in each of the $N_{\text{obs}}$ estimated GW distances equal to
\begin{equation}
\sigma_i = \sqrt{\sigma_{\text{ins}}(z_i, \bm{s}_i)^2 + \sigma_{\text{lens}}(z_i)^2}\,, \label{sigma}
\end{equation}
for $ i = 1, ..., N_{\text{obs}}$. Here the errors will always be evaluated at the fiducial model employed in the simulations.

Given a fiducial gravitational and cosmological model, we draw the redshift of the i-th binary source from the distribution
\begin{equation}
\rho (z_i) = N_{z} \frac{4\pi [D_c(z_i)]^2}{(1 + z_i)^2 \mathcal{H}(z_i)} r(z_i)\,, \label{z_pdf}
\end{equation}
where $D_c(z) = \int_0^z dz/H(z)$ is the fiducial comoving distance, $r(z)$ is the rate of BNS mergers evolution and $N_z$ is a normalization factor. For the parameters $\bm{s}$, the distributions $\rho(\bm{s})$ assumed in this work are the following. The masses were uniformly sampled from [1, 2]$M_{\odot}$; $\iota$ was sampled from [\ang{0}, \ang{20}] to be consistent with the detection of a gamma ray burst \cite{Zhao2011}; for $(\theta, \phi)$ we made a uniform sampling in the sky and $\psi$ was drawn from [0, 2$\pi$].

Finally, the GW distance $D^{\textrm{gw}}_i$ of the $i$-th event is drawn from a Gaussian distribution, with standard deviation given by Eq.\ (\ref{sigma}), centered in its theoretical value $D^{\textrm{gw}}(z_i, \bm{\Theta})$, where $\bm{\Theta}$ are the cosmological and gravitational parameters defining the model assumed in Eq.\ \eqref{gw_distance}. One may obtain, then, the posterior probability for the parameters of interest, given the simulated data. That is
\begin{equation}
\rho(\bm{\Theta}| \bm{D}^{\text{gw}} ) \propto  \rho (\bm{\Theta}) \exp \Bigg\{-\sum_{i = 1}^{N_{\text{obs}}} \frac{\left[D^{\text{gw}}_{i} - D^{\text{gw}}(z_i, \bm{\Theta})\right]^2}{2\sigma_i^2}\Bigg\}\,. \label{final_posterior}
\end{equation}

We consider reliable only those events with SNR $>$ 8 and discard the others as in \cite{Matos2021, DAgostino2019}, which deforms the ultimate probability density of redshifts, as expected. Errors in the inference of the redshifts from the EM signals, expected to be subdominant, were not included in our analysis, neither selection effects that depend on the network of telescopes that will be operating at the same time as ET. Also, other sources of uncertainty were neglected in this work, such as overdensities near emission and peculiar velocities. The latter should be important only at low redshifts, where the population of mergers is of little relevance.

In our analysis, no dependency on the redshift distribution of the binaries was added to the above likelihood. This distribution depends in general on the star formation rate and it is sensitive to the cosmological background $\mathcal{H}(\tau)$, affecting the constraints on the cosmological parameters and the dark energy equation of state, particularly when doing forecasts where also the total amount of events is an unknown variable. However, here our main interest is on the constraining power of standard sirens on the modifications of gravity described in Section \ref{sec:theory}, which are encoded in the distance-redshift relation. Thus, we instead compute only the probability density of $\bm{\Theta}$ given the fact that, for a binary source at redshift $z_i$, precisely obtained via its EM counterpart, one measures a distance $D^{\text{gw}}_i$ from the GW signal, regardless of how the sources of such signals are distributed in the sky.

\subsection{Planck}

Another data set included in our analysis consists of the temperature and polarization maps of the CMB measured by the \textit{Planck} mission in 2018. Each Planck likelihood is described in detail in \cite{Planck2018V, Planck2018VIII}, and so here we simply summarize our notation:
\begin{itemize}
    \item \textit{Planck} TT,TE,EE is the product of two likelihoods: the combination of the high-multipole ($\ell \geq 30$) $\mathtt{Plik}$ likelihoods from temperature (TT), E-mode polarization (EE), and temperature-polarization (TE) power spectra, including their correlations; and the low-multipole ($2 \leq \ell < 30$) $\mathtt{Commander}$ likelihood for the temperature-only power spectrum;
    
    \item \textit{Planck} lowE is the $\mathtt{SimAll}$ likelihood from the low-multipole ($2 \leq \ell < 30$) E-mode power spectrum;
    
    \item \textit{Planck} lensing is the likelihood from the CMB lensing power spectrum.
\end{itemize}

We refer for brevity to the joint likelihoods from the temperature and polarization power spectra, i.e. \textit{Planck} TT,TE,EE + lowE, as ``CMB" and its combination with {\it Planck} lensing as ``CMB+lens".

\subsection{Large-scale structure likelihoods}

Large-scale structure surveys, particularly the combination of galaxy number counts and weak lensing surveys, are well suited to constrain the gravitational slip \cite{Daniel2008}. 
Here we use as a particular example a 15'000 square degree survey that will measure number counts and galaxy shapes out to a redshift of about 2, modeled on the Euclid specifications. Specifically, we use the likelihoods from \cite{Sprenger2019}. These likelihoods are in turn based on an older version described in \cite{Audren2013}. We briefly describe them below.

The \textit{Euclid-like weak lensing} likelihood (called WL in the rest of this paper) is based on measurements of cosmic shear, that is, the alignment of galaxies generated by weak gravitational lensing due to the structure present in the light path between the galaxy and the observer. The correlations between such alignments are computed at 10 different redshift bins, each with 30 galaxies per square-arcmin. The resulting shear power spectrum coefficients $C^{ij}_{\ell}$ $(i,j = 1, \dots, 10)$ can be related to the matter power spectrum $P(k,z)$ via
\begin{equation}
C^{ij}_{\ell} = \frac{9}{16}\Omega_{m0}H_0^4\int_0^{\infty}\frac{dr}{r^2}g_i(r)g_j(r)\Sigma\big(\ell/r, z(r)\big)^2P\big(\ell/r,z(r)\big)\,,
\end{equation}
where $g_i$ is a function of the distribution of galaxies in the i-th redshift bin. A systematic error due to intrinsic alignments is then added to $C^{ij}_{\ell}$. We highlight that in addition to the effect of the modification of gravity inside $P(k,z)$, we have to account explicitly for the extra $\Sigma^2$ factor in view of eq. (\ref{lensing}).

The \textit{Euclid-like P(k)} likelihood (GC) quantifies the probability of the cosmological model given the distribution of galaxies in the sky in redshift space. The relation between the observable and the theoretical isotropic matter power spectrum is computed considering the Alcock-Paczy\'nski effect, the limited resolution of the instruments and redshift-space distortions. The RSD modelling includes both the Kaiser formula, concerning the large scales, and the distortions in the small scales due to the peculiar velocities of galaxies. Furthermore, a linear and scale-independent galaxy bias is assumed. Contrary to the weak lensing \euclid-like likelihood, where we inserted the $\Sigma^2$ correction, for \euclid-like $P(k)$ no modification in the likelihood already implemented in \texttt{MontePython} is necessary.

\section{Results}
\label{sec:results}

In this section we present the results of the parameter estimation from our Monte Carlo Markov Chains (MCMCs). They were generated using version 3.5.0 of the $\mathtt{MontePython}$ code \cite{Audren2012wb, Brinckmann2018cvx}, with the help of $\mathtt{hi\_class}$ \cite{Zumalacarregui2017, Bellini2020}, a modification of the Einstein-Boltzmann solver $\mathtt{CLASS}$ \cite{Blas2011} that computes cosmological observables in the wider context of the Horndeski family of gravity models.

\subsection{Slip \textit{versus} GWs} \label{subsec:slip_vs_gws}

We are interested to investigate first to which precision one can infer $\Omega_0$ whenever the slip is measured at a certain accuracy level. For this purpose, we assume a Gaussian likelihood for $\eta$ at a certain redshift $z$ and scale $k$ with standard deviation of 1\% around the $\Lambda$CDM value of 1 and fix all the cosmological parameters to the values in Table \ref{tab:fiducial_model}. Notice that Eq.\ \eqref{phi_minus_psi} is scale dependent being subject to horizon effects, and so in practice we have to specify a scale. By assuming that the slip is measured today ($z=0$) at either $k = 0.1 \; \text{Mpc}^{-1}$ or $k = 0.01 \; \text{Mpc}^{-1}$, the 68\% CL found for $\Omega_0$ was
\begin{align}
 \Delta \Omega_0 = 0.015
\end{align}
for both scales.
Since the error does not depend on the choice of scale for the two cases considered, as expected from the weak $k$-dependence of the slip inside the horizon (cf. figure \ref{fig:xi}), we will only consider the larger scale from now on.

In our baseline model, the Planck constant changes linearly with the scale factor, so that the gravitational slip is largest at late times (cf.\ Fig.\ \ref{fig:xi}). It is then no surprise that measuring the slip with the same precision at higher redshift, $z=1.1$, results in a significantly larger uncertainty, 
\begin{align}
 \Delta \Omega_0 = 0.07
\end{align}
(for $k=0.01 \; \text{Mpc}^{-1}$).

We then considered a more realistic scenario in which the slip is assumed to be measured at $k = 0.01 \; \text{Mpc}^{-1}$ in three redshift bins, according to the quite model-independent forecasts in Table X of \cite{Amendola2014}: at $z = 0.7$, $\Delta \eta = 3.1\%$, at $z = 1.1$, $\Delta \eta = 3.7\%$ and at $z = 1.5$, $\Delta \eta = 3.2\%$. The result is
\begin{align}
 \Delta \Omega_0 = 0.12\,.
\end{align}

Furthermore, we also simulated a data set of GW events detected by the Einstein Telescope with electromagnetic counterpart. Again, by fixing all other parameters, we get the following 68\% CL on $\Omega_0$ for two scenarios:
\begin{align}
\Delta \Omega_0 =
\begin{cases}
0.08\,, \quad \,\,\, \text{100 GWs}\\
0.035\,, \quad \text{500 GWs}.
\end{cases}
\end{align}
The improvement in the error when increasing the number of GW events is consistent with the expected $\sqrt{N}$ scaling. 

\begin{figure}
    \centering
    \includegraphics[scale=0.5]{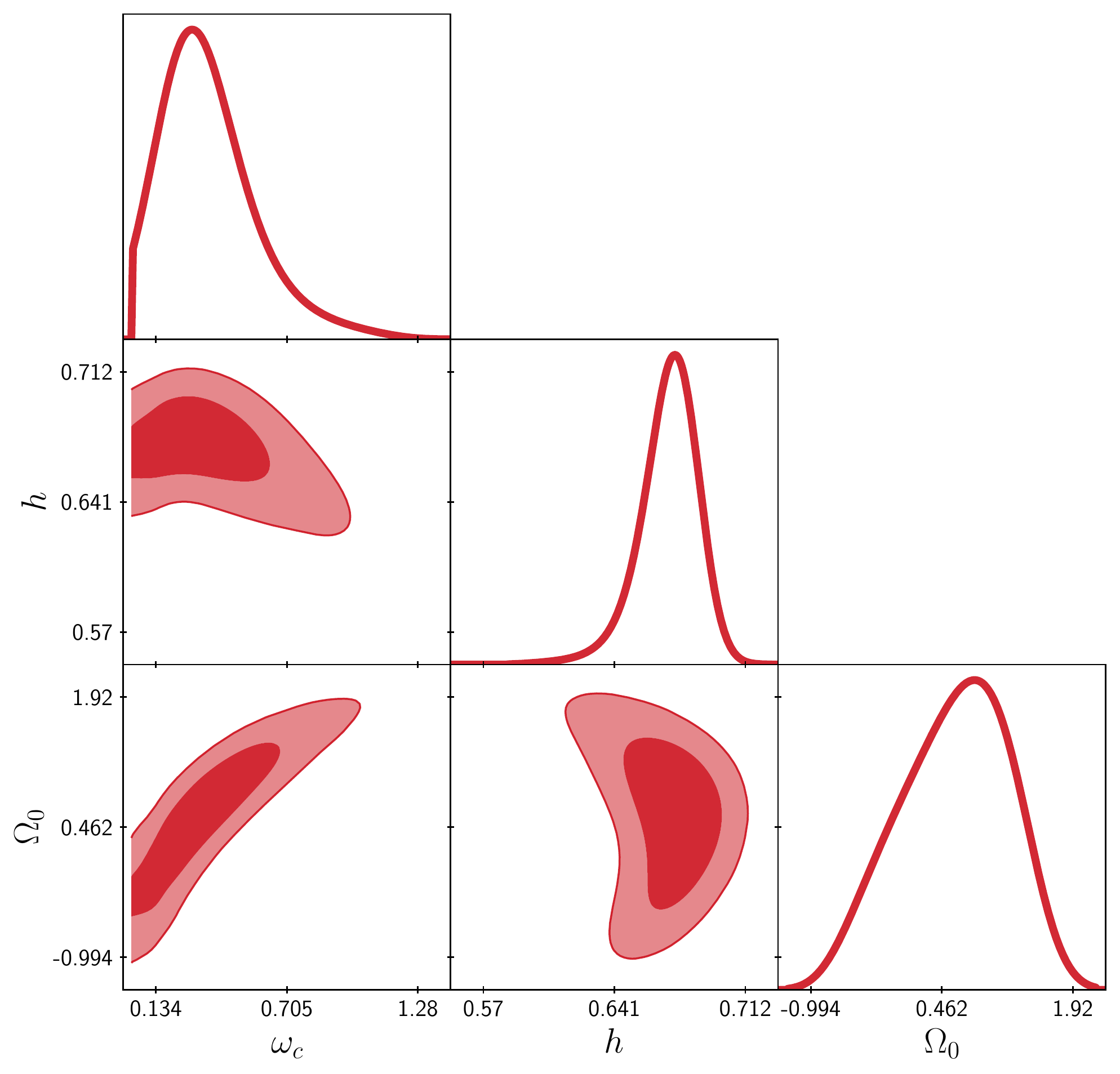}
    \caption{Constraints from five hundred multimessenger gravitational wave events as forecasted for the Einstein Telescope.}
    \label{fig:gw_500_3d}
\end{figure}

The case with 500 GW events described above provides a measurement of the deviation from GR at the percent level of accuracy. This is our best-case scenario though still realistic, once we consider that the cosmological parameters present in the GW likelihood, $h$ and $\omega_{c}$, are independently measured with high precision by other probes. On the other hand, the less informative case occurs when letting the cosmological parameters also free still only with GW mock data. The resulting contours in this case are shown in Figure \ref{fig:gw_500_3d}.

\begin{figure}
    \centering
    \includegraphics[scale=0.6, trim = 1.1cm 0 0 0]{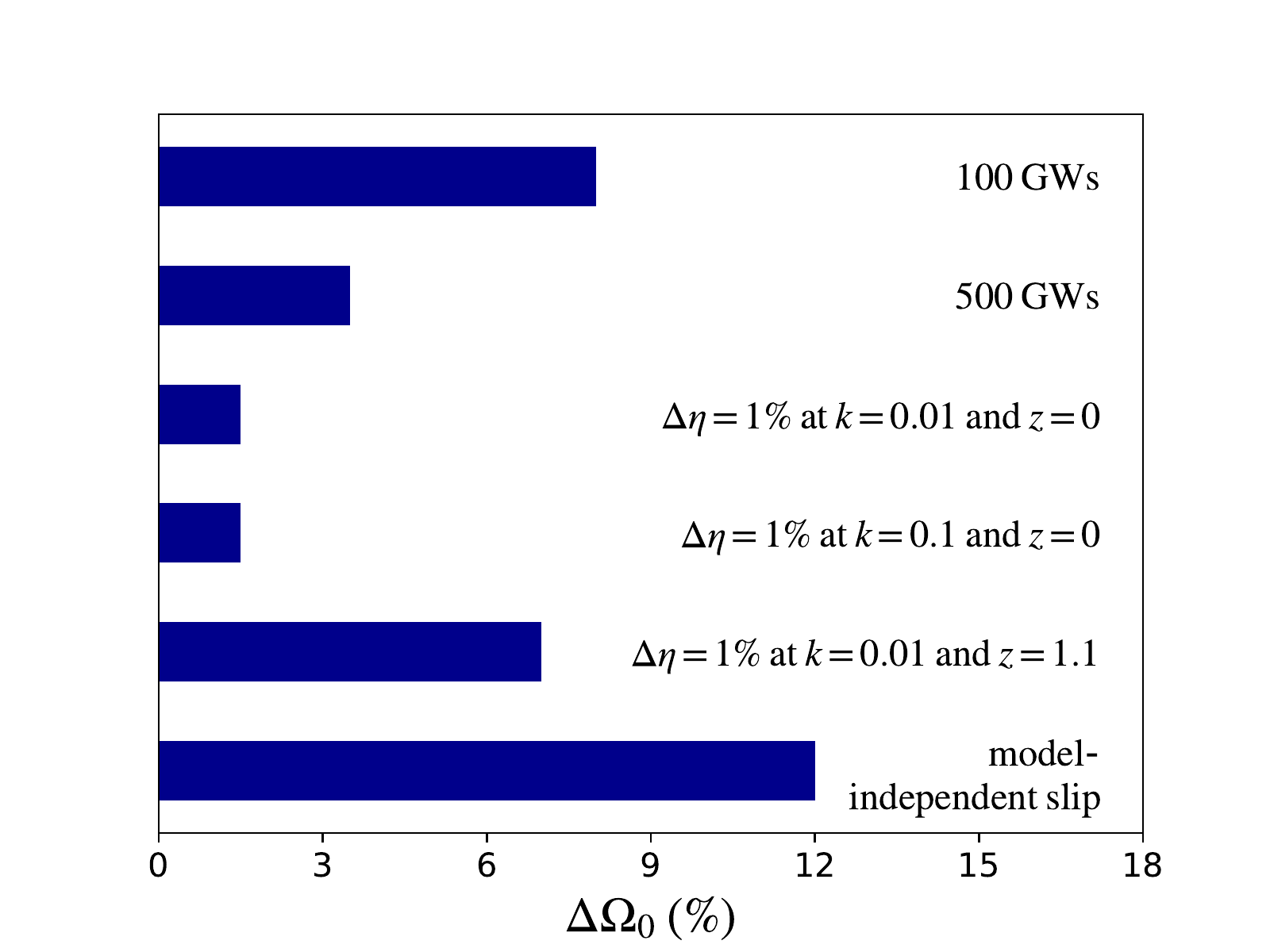}
    \caption{Comparison between the 68\% CLs for $\Omega_0$ from hypothetical measurements of the slip parameter or GW detections.}
    \label{fig:fake_slip_gws}
\end{figure}

As a summary, Figure \ref{fig:fake_slip_gws} shows a comparison between these constraints. We point out that, from these first simple estimates, depending on the modelling of the slip measurement and the exact number of observed multimessenger GW events, one or the other might be a better probe of the deviations from GR that we study here. This justifies performing the more robust analysis with actual forecasts on the slip side that follows next. The values of the marginalized one dimensional 68\% CLs for the main cases studied are summarized in Table \ref{tab:results}.

\begin{table*}[t] 
	\centering
	
	\setlength{\extrarowheight}{10pt}
	
\begin{tabularx}{\textwidth} { 
   >{\raggedright\arraybackslash}X 
   >{\centering\arraybackslash}X 
   >{\centering\arraybackslash}X >{\centering\arraybackslash}X >{\centering\arraybackslash}X  }
  \hhline{= = = = =}
		Parameter & WL+GC & CMB+lens & GW & All data\\ \hline
	    $\omega_{c}$ \dotfill & $\pm 0.0006$ & $\pm 0.0009$ & $^{+0.13}_{-0.27}$ & $\pm 0.0003$\\
	    
		$10^9 A_s$ \dotfill & $\pm 0.035$ & $\pm 0.012$ & $-$ & $\pm 0.01$\\
		
		$h$ \dotfill & $\pm 0.0023$ & $\pm 0.0033$ & $^{+0.017}_{-0.013}$ & $\pm 0.001$\\ 
		$\Omega_0$ \dotfill & $\pm 0.01$  & $\pm 0.026$ & $^{+0.7}_{-0.5}$ & $\pm 0.006$ \\[2mm]  \hline
		\end{tabularx}
		\caption{Marginalized one-dimensional 68\% CL errors on the cosmological parameters and $\Omega_0$ for the analysed data sets and their combination. The last column shows the results of the combination of all datasets considered, i.e. CMB+lens + WL+GC + GW.}
		\label{tab:results}
\end{table*}

\subsection{Planck}

\begin{figure*} 
    \centering
    \includegraphics[scale=0.4]{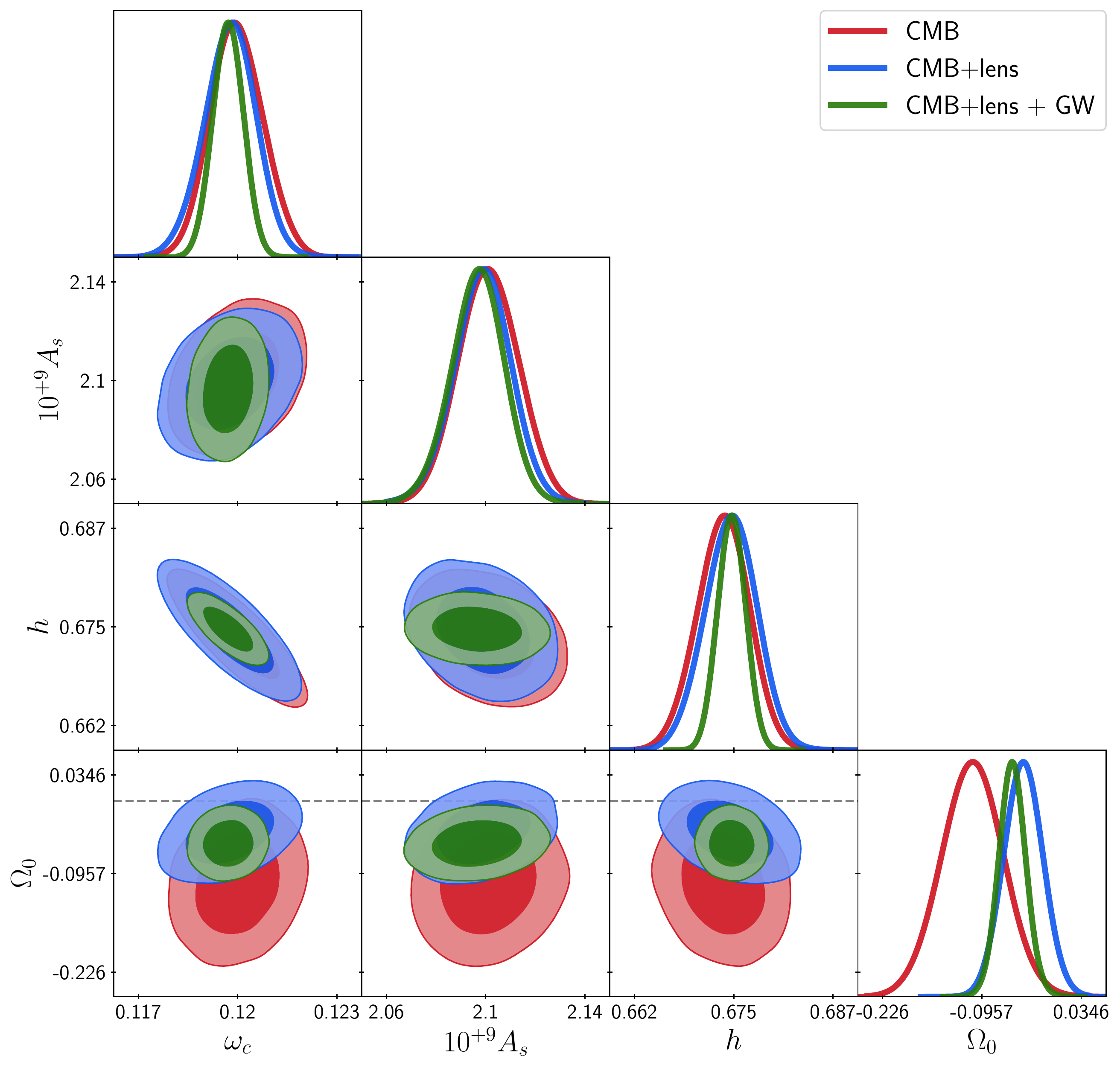}
    \caption{Constraints from Planck TT,TE,EE+lowE (red), its combination with Planck lensing and the further combination with the forecasted five hundred GWs from the Einstein Telescope.}
    \label{fig:planck_results}
\end{figure*}

We now present the results from current CMB data and its combination with GWs. Figure \ref{fig:planck_results} shows the two-dimensional marginalized 68\% CLs in the parameter space of our base model for three different cases:  \textit{Planck} TT,TE,EE+lowE, its combination with \textit{Planck} lensing and their further combination with five hundred GWs from the Einstein Telescope. The main result to be stressed here is that, in all cases, we get a percent level accuracy on $\Omega_0$, which is in agreement with a similar analysis performed in \cite{Planck2018} for a different parametrization of $\alpha_M$ and all cosmological parameters free. 

\begin{figure}
    \centering
    \includegraphics[scale=0.55, trim= 0 0 0 0]{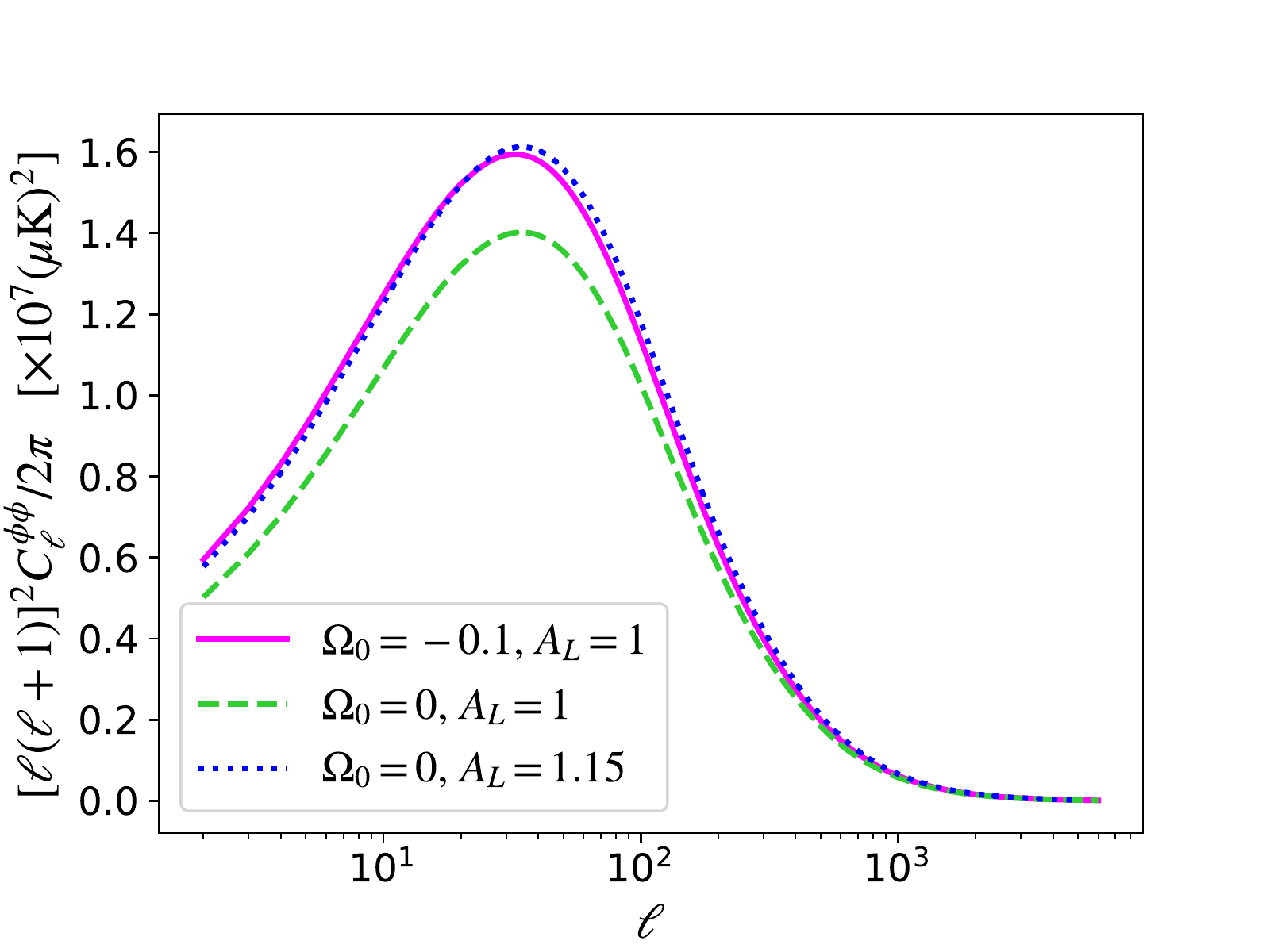}
    \caption{Comparison between the lensing potential power spectrum coefficients for the following models: MG with $\Omega_0 = -0.1$ and $A_L = 1$, $\Lambda$CDM with $A_L = 1.15$ and $\Lambda$CDM with $A_L = 1$. All other cosmological parameters have the same values.}
    \label{fig:lensing_cls}
\end{figure}

We also 	point out that \textit{Planck} favours MG models with a decreasing Planck mass (in time) and, furthermore, that \textit{Planck} TT,TE,EE+lowE alone recovers the $\Lambda$CDM value $\Omega_0 = 0$ only at $2\sigma$ from the best fit model ($\Omega_0 = -0.1)$. This was already noticed and discussed in, for instance, \cite{Planck2015DE} (see Figure 17) and \cite{Planck2015CP} (subsection 5.1), where the authors attribute this feature to a degeneracy between $(\mu, \Sigma)$ and the lensing amplitude calibration $A_L$. In fact, Figure \ref{fig:lensing_cls} shows that adding a small variation to the Planck mass has the same effect in the lensing potential power spectrum as allowing for an overall amplitude slightly different from 1; we checked that when letting $A_L$ free the error in $\Omega_0$ is roughly multiplied by a factor 3. When adding the \textit{Planck} lensing likelihood, as we see again in Figure \ref{fig:planck_results}, the contours are pushed towards GR $\Lambda$CDM (dashed grey line). We comment that, on the other hand, more updated CMB analysis alleviate or even do not show this preference for larger values of $A_L$ \cite{Aiola2020} and therefore one should check the consistency of the bounds on $\Omega_0$ across different CMB likelihoods.

Regarding the combination of CMB and GWs, by also looking at Figure \ref{fig:gw_500_3d} we note that CMB helps setting the matter density parameter $\omega_{c}$ that is poorly inferred from GWs, while the addition of data in the tensor sector improves the measurement of $\Omega_0$ reducing its error by 40\%. While this is a significant improvement, we stress that the more important aspect of testing the distance-redshift relation with future GW data comes from its ability to probe one of the MG property functions ($\alpha_M$) alone, while CMB and LSS in general are sensitive to combinations of all the $\alpha$'s. This advantage cannot be illustrated with our simple base model, since it  summarizes all the deviations from GR in only one degree of freedom.

\begin{table*}[t] 
	\centering
	
	\setlength{\extrarowheight}{10pt}
	
\newcolumntype{s}{>{\hsize=.6\hsize \centering\arraybackslash}X}
\newcolumntype{l}{>{\hsize=\hsize \centering\arraybackslash}X}

\begin{tabularx}{\textwidth} { 
   >{\hsize=.6\hsize\raggedright\arraybackslash}X 
   ssllslls  }
  \hhline{= = = = = = = = =}
		 & GW & WL & WL+GC & WL+GC + GW & CMB & CMB+lens  & CMB+lens + GW & All data \\ \hline
		$\Delta\Omega_0$ & $^{+0.7}_{-0.5}$ & $\pm 0.1$ & $\pm 0.01$ & $\pm 0.01$ & $\pm 0.042$ & $\pm 0.026$ & $\pm 0.02$ & $\pm 0.006$ \\[2mm] \hline
		\end{tabularx}
		\caption{Marginalized one-dimensional 68\% CL errors on $\Omega_0$ for all the analysed data sets and their combinations. The last column shows the results of the combination of all datasets considered, i.e. CMB+lens +WL+GC + GW.}
		\label{tab:results_Omega_0}
\end{table*}

\subsection{Large-scale structure data} \label{subsec:euclid}

\begin{figure} 
    \centering
    \includegraphics[scale=0.35]{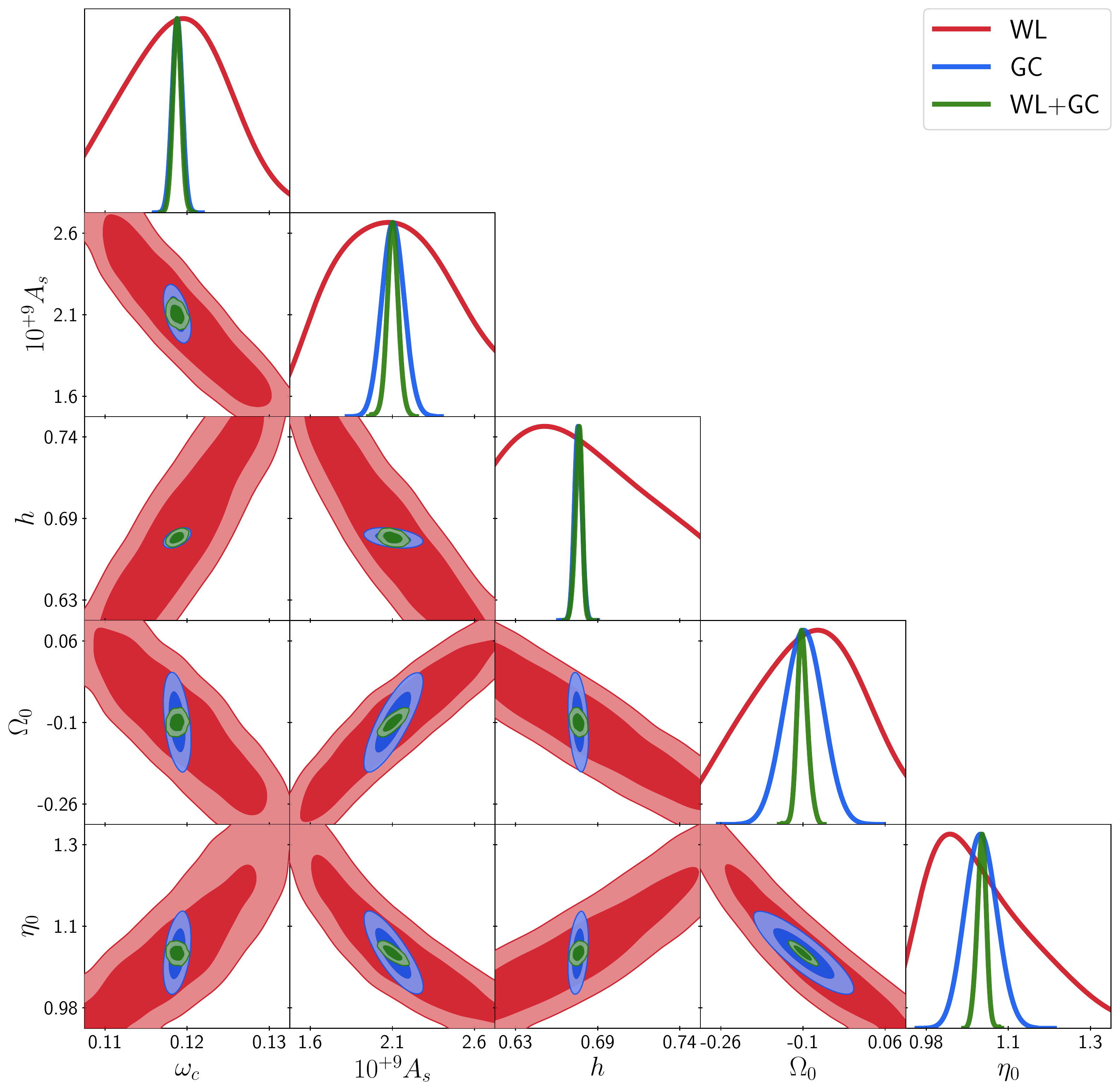}
    \caption{Forecasted constraints from weak lensing (red), galaxy clustering (blue) and their combination (green) for a \euclid-like survey. The further combination with five hundred multimessenger GW detections by the Einstein Telescope was omitted since it generates similar contours as the green ones.}
\label{fig:euclid_results}
\end{figure}

We now present the results from \euclid-like forecasts. Figure \ref{fig:euclid_results} shows the two dimensional marginalized 68\% and 95\% CLs on the parameter space of our models for, in red, weak lensing with uniform priors, in blue, galaxy clustering and their combination. Here we also show the result for the derived parameter $\eta_0$ which is the value of the gravitational slip today at $k=0.01 \; \textrm{Mpc}^{-1}$. We remind that our fiducial model features a 10\% deviation from 1 on the Planck mass and that it coincides with the bestfit model of the \textit{Planck} primary spectrum. We conclude that such model would be easily distinguished from GR's $\Lambda$CDM by a near-future \euclid-like large scale structure survey alone in $\sim 10 \sigma$, through measurements of both $\eta_0$ and $\Omega_0$ with 1\% of accuracy.

\begin{figure} 
    \centering
    \includegraphics[scale=0.4]{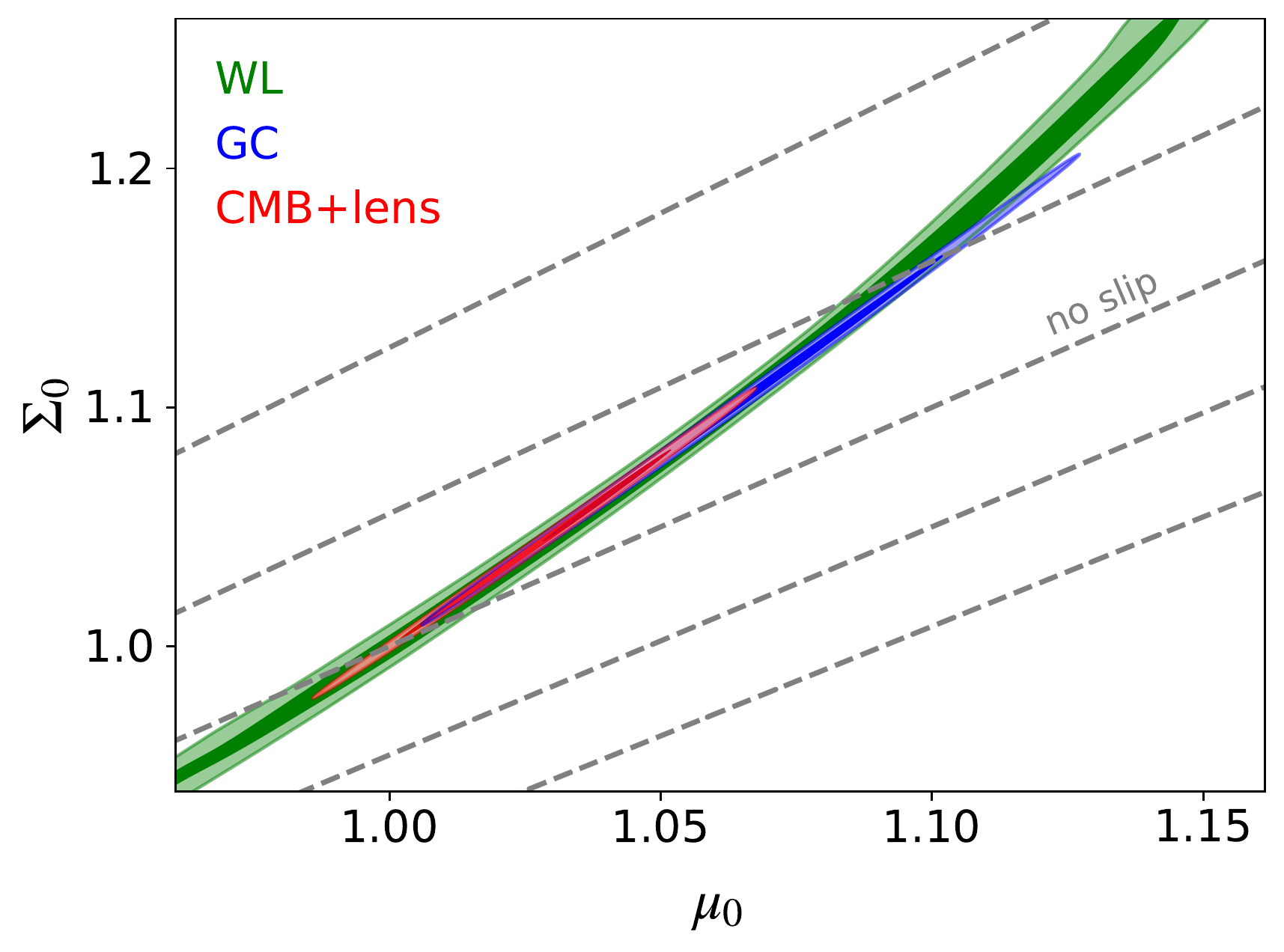}
    \caption{Constraints from large-scale structure on the derived parameter space ($\mu$, $\Sigma$), where the sub-index 0 stands for evaluation today and $k = 0.01$/Mpc. Dashed grey lines correspond to loci of constant gravitational slip ($\delta \eta = 0.1$ between consecutive lines).}
\label{fig:mu_Sigma}
\end{figure}

We comment that, in the approach where the functions $\mu$ and $\Sigma$ are independently parametrized, it is expected that weak lensing and galaxy clustering data would be complementary to infer the slip, since each of them probe different combinations of the scalar potentials. However, in our analysis we find that WL provides much weaker constraints than GC, even though it is still relevant for constraining the MG deviations, as we can see in figure \ref{fig:euclid_results}.  
This is due to the fact that, again, the models here considered have only one extra degree of freedom $\alpha_M$, which is exactly what allows us to connect GW and slip measurements without further assumptions. This creates a relation between the derived $\mu$ and $\Sigma$, depending on the cosmological parameters, as illustrated in figure \ref{fig:mu_Sigma}, by noticing how their values (at a certain scale and redshift) are correlated for the different analysed data sets.

Finally, we find that the addition of five hundred GWs from the Einstein Telescope alone, despite constituting an independent test, does not significantly improve the results from our large-scale structure likelihood. We leave the comparison between the outcome of such tests in the scalar and tensor sectors for the next subsection.

\subsection{Combination of data sets}

\begin{figure*} 
    \centering
    \includegraphics[scale=0.4]{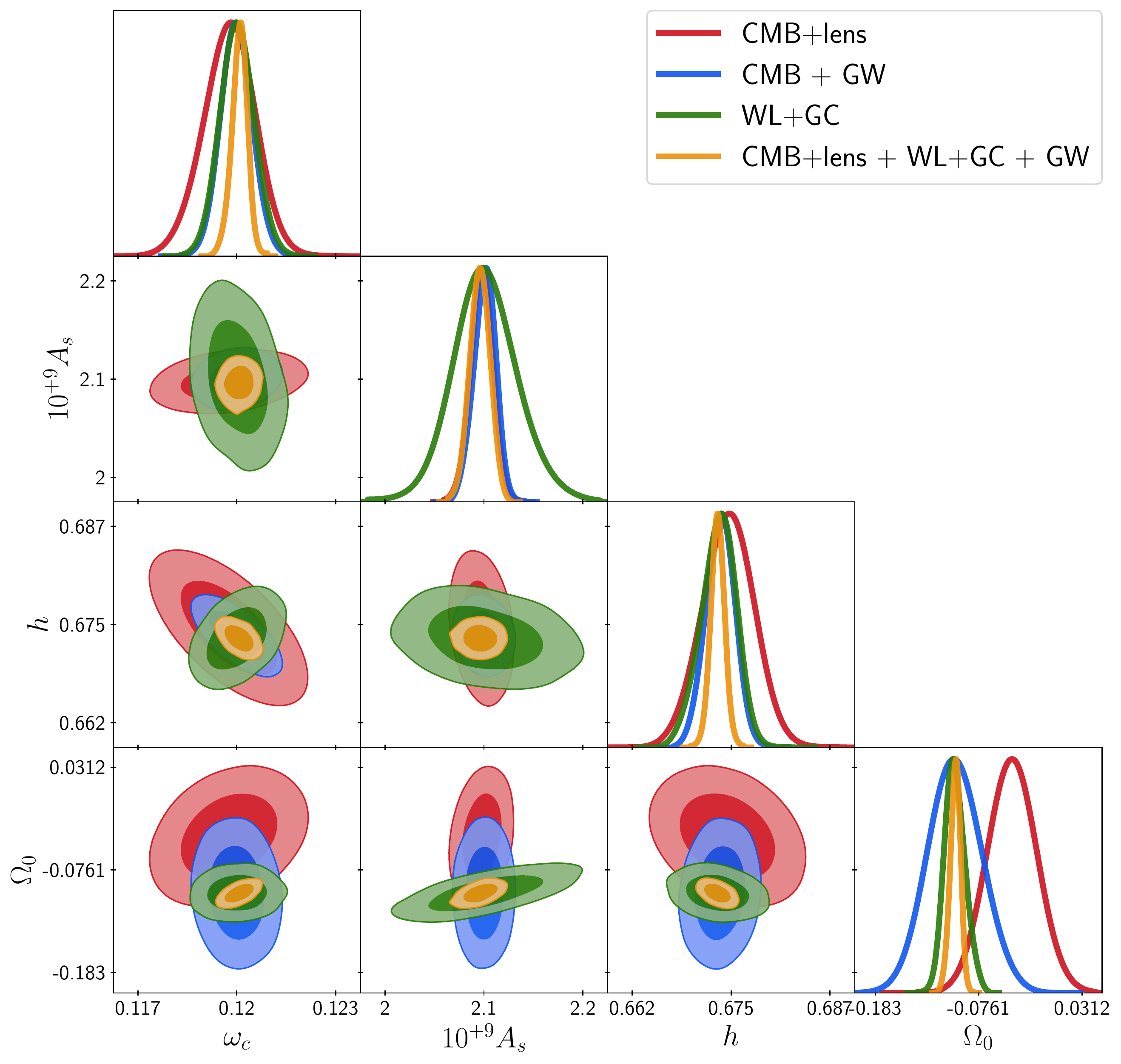}
    \caption{Comparison between the constraints from \textit{Planck} TT,TE,EE+lowE+lensing; five hundred GWs from the Einstein Telescope combined with \textit{Planck} primary spectra; from weak lensing and galaxy clustering; and the final combination of all data sets.}\label{fig:combined_results}
\end{figure*}

We finally compare the constraining power of our different data sets, as summarized in figure \ref{fig:combined_results}. First, by comparing the green and red 68\% and 95\% contours, we see that a \euclid-like survey and \textit{Planck} give complementary information on the cosmological parameters. The former, however, provides a measurement of the MG parameter $\Omega_0$ twice more accurate than that of current CMB data (see also table \ref{tab:results}).

As mentioned in the last subsection, we obtain that a \euclid-like large-scale structure survey alone will measure the slip with 1\% error. Then, one might wish to compare its corresponding bound on $\Omega_0$ (of also 1\%, see Table \ref{tab:results_Omega_0}) with the results of our rough estimation in subsection \ref{subsec:slip_vs_gws}, where we fixed all cosmological parameters and assumed a hypothetical 1\% accuracy for the slip. Since GWs alone do not constrain the cosmological parameters due to large degeneracies, we can use \textit{Planck} TT,TE,EE+lowE to fix them, obtaining the blue contours in figure \ref{fig:combined_results}; one gets then a 3\% error in $\Omega_0$. From these we conclude that future measurements of the slip from \euclid-like surveys, for the simple gravity models here studied, will reach slightly higher accuracy levels in measurements of deviations of GR via $\alpha_M$, as compared to future GW data from the Einstein Telescope, when additional data (like from the CMB) constrains the other cosmological parameters sufficiently.

Futhermore, the addition of future data on both sectors, here data from large-scale structure and multimessenger GW events, will substantially improve the current results of \textit{Planck}, as we notice by comparing the yellow and red contours in Figure \ref{fig:combined_results}. More precisely, by looking at Table \ref{tab:results}, we see that the errors in the background cosmological parameters $\omega_c$ and $h$ and the MG parameter $\Omega_0$ are decreased in around 67\%, 70\% and 77\%, respectively.

{Finally, we obtain that the bounds on the derived parameters $\mu_0$ and $\Sigma_0$ from the combination CMB+lens+WL+GC+GW will reach the sub-percent level, as well as the slip, with $\Delta\eta_0 = 0.5\%$. However, we remind that since our MG models have only one extra degree of freedom and $\mu$ and $\Sigma$ are related, they end up being more strongly constrained than in a scenario where they are independently parametrized, and thus these strong bounds are actually driven by the choice of model.}

\subsection{No-slip model}

We also studied the constraints from the three data sets, \textit{Planck}, standard sirens and large-scale structure, on the no-slip model described in subsection \ref{subsec:no_slip}. In the analysis of \cite{Linder2018}, a no-slip model with $A = 0.03$ is compatible with measurements of the growth rate $f\sigma_8$. Here, however, even if it generates deviations in the CMB spectra, when compared to $\Lambda$CDM, of the same order as our base model, it turns out to be completely excluded by Planck data. This is because \planck\ favors models with decreasing Planck mass, as discussed before, and so it constrains the viable no-slip models better than our base-models with negative $\Omega_0$, since they have increasing Planck mass.

\begin{figure*} 
    \centering
    \includegraphics[scale=0.35]{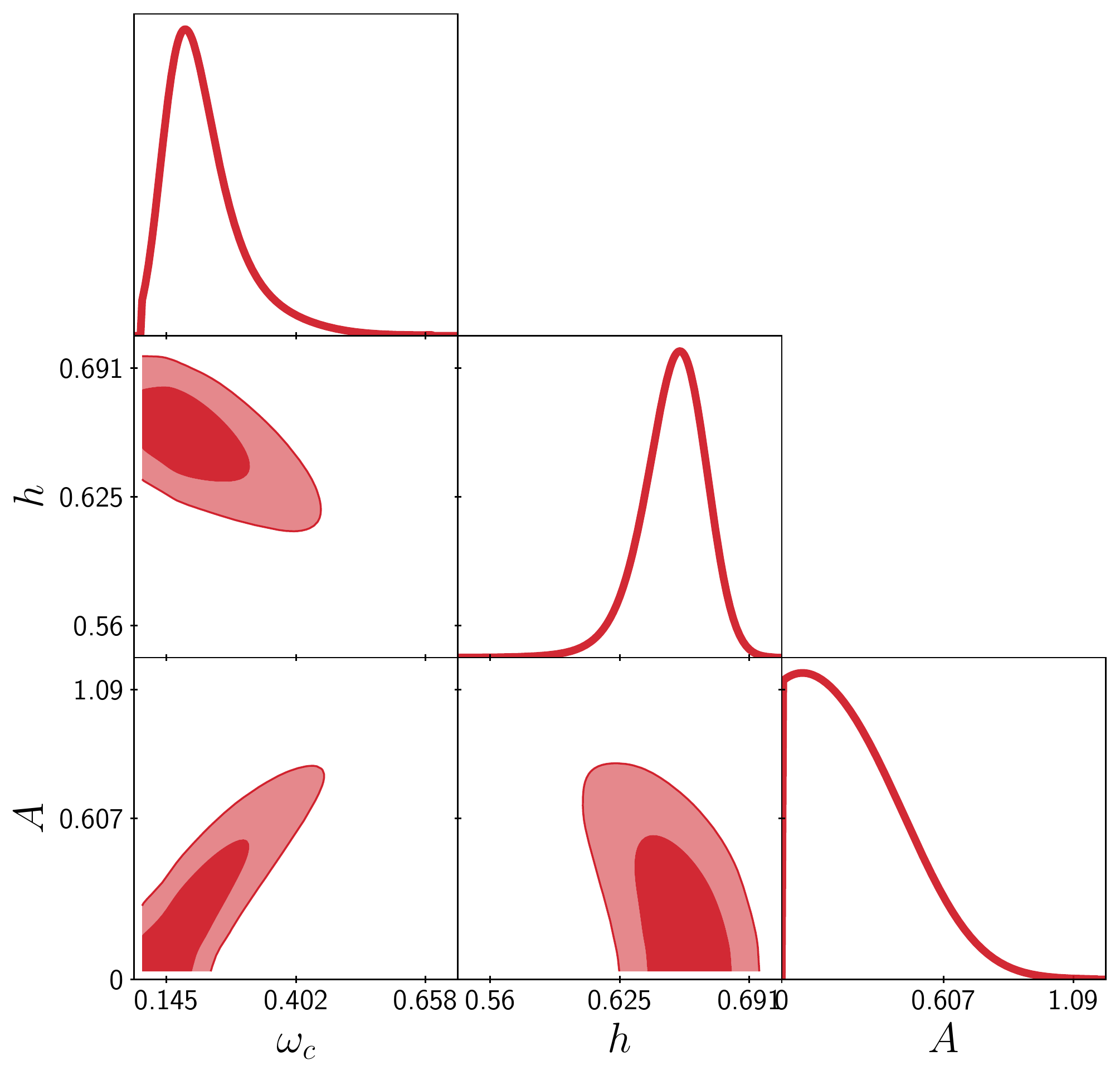}
    \includegraphics[scale=0.35]{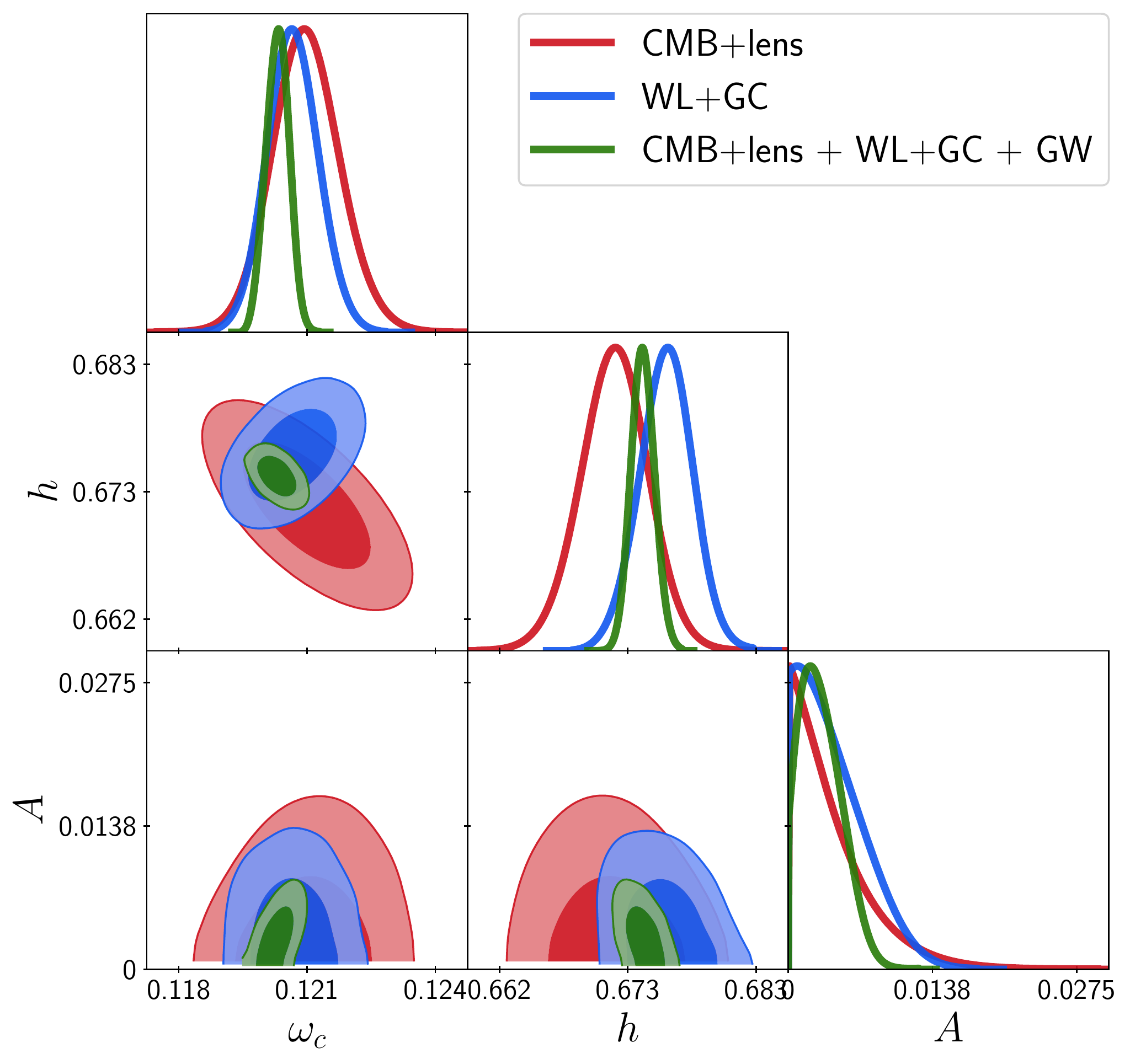}
    \caption{Results for viable no-slip models marginalized in $A_s$. On the left, 68\% and 95\% CLs. from five hundred standard sirens with the Einstein Telescope and, on the right, Planck TT,TE,EE+lowE+lensing, galaxy clustering and weak lensing, and  their combination with five hundred GWs.} \label{fig:no_slip_results}
\end{figure*}

In the \euclid-like and Einstein Telescope forecasts we used for the fiducial model the values of the cosmological parameters given in Table \ref{tab:fiducial_model} and $A = 0.005$, which is compatible with \textit{Planck} data at 1$\sigma$. We also imposed the prior $A>0$ in order to avoid instability issues as discussed previously. Figure \ref{fig:no_slip_results} shows the constraints from GWs, CMB, LSS and their combinations assuming the Universe is well described by a no-slip model as far as possible from $\Lambda$CDM while still being compatible with CMB data.

Table \ref{tab:results_no_slip} shows the 1-dimensional bounds on the cosmological parameters and $A$. We comment that $\Lambda$CDM ($A = 0$) is compatible within $1\sigma$ with all data sets and their combinations. {We also notice in Fig.\ \ref{fig:no_slip_results} that $A$ and $\omega_c$ are highly degenerate, which prevents GW data alone from putting strong bounds on the deviation from GR, as discussed in \cite{Mitra2021}, where they use a parametrization for $\Xi$ that does not consider the viability conditions of no-slip gravity.}

Remarkably, these models have vanishing slip at late times, and thus purely (model-independent) slip measurements in principle would not be able to constrain them. However, here we conclude that LSS still gives better bounds on the deviation of the Planck mass for the analysed no-slip models, and this is because both weak lensing and galaxy clustering are separately modified.

\begin{table*}[t] 
	\centering
	
	\setlength{\extrarowheight}{10pt}
	
\begin{tabularx}{\textwidth} { 
   >{\raggedright\arraybackslash}X 
   >{\centering\arraybackslash}X 
   >{\centering\arraybackslash}X >{\centering\arraybackslash}X >{\centering\arraybackslash}X  }
  \hhline{= = = = =}
		Parameter & WL+GC & CMB+lens & GW & All data\\ \hline
	    $\omega_{c}$ \dotfill & $\pm 0.0006$ & $\pm 0.0008$ & $^{+0.03}_{-0.10}$ & $^{+0.0002}_{-0.0003}$\\
	    
		$10^9 A_s$ \dotfill & $\pm 0.03$ & $\pm 0.011$ & $-$ & $\pm 0.01$\\
		
		$h$ \dotfill & $\pm 0.002$ & $\pm 0.003$ & $^{+0.018}_{-0.013}$ & $\pm 0.001$\\
		
		$A$ \dotfill & $< 0.005 \; (0.006)$  & $< 0.005 \; (0.006)$ & $< 0.36 \; (0.44)$ & $< 0.004 \; (0.006)$ \\[2mm]  \hline
		\end{tabularx}
		\caption{Marginalized one-dimensional 68\% CL errors on the parameters of the no-slip model for the analysed data sets and their combination (CMB+lens + WL+GC + GW). The last row shows the 68\% (95\%) upper bounds on $A$, the deviation of $\Lambda$CDM.}
		\label{tab:results_no_slip}
\end{table*}

\section{Conclusion}
\label{sec:conclusion}

The starting point of our investigation was the question how measurements of the gravitational slip from large-scale structure and of GW distances compare when the goal is to constrain an underlying MG theory. Often, the GW side and the LSS side are considered separately. This is particularly the case when using so-called `phenomenological modified gravity' parameters like $\mu$, $\Sigma$ and $\Xi$. However, for a given action-based model, those parameters are generally linked to the fundamental model parameters and so are not independent. Here we illustrate this with the help of effective field theory, where a running of the Planck mass affects both the GW friction and the gravitational slip simultaneously.

As a first step, we use an order-of-magnitude estimate for the gravitational slip constraints, and compare the corresponding limits on the non-minimal coupling parameter $\Omega_0$ with those from predictions for the Einstein Telescope. Fig.\ \ref{fig:fake_slip_gws} shows that, depending on the redshift at which we infer the slip or the exact amount of GW events, one data set might be better than the other, so it makes sense to perform a more precise investigation.

Comparing CMB observations from the {\it Planck} satellite and galaxy clustering as well as weak lensing data from a survey like the upcoming \euclid\ satellite with GW distances from the Einstein Telescope, we find that generally the large-scale structure data constrains the Planck mass run rate more strongly, see Fig.~\ref{fig:combined_results} {and Table \ref{tab:results_Omega_0}}. We conclude that we get stronger bounds on the Planck mass from the scalar sector than from the tensor sector for the particular class of models featuring only this extra degree of freedom, described in Subsection \ref{subsec:base_model}. However, it is important to emphasize a few points: Firstly, in models where many parameters are varying (rather than a single one as in our investigation) we will need multiple observations to constrain all the parameters {simultaneously}. In this case the inclusion of standard sirens might be crucial since they remain sensitive to only $\alpha_M$ (and this is why our results from the tensor sector are less model-dependent). Secondly, GW and LSS data do measure fundamentally different observables, and they can only be compared within the context of a particular model. Clearly, both types of observations are fundamentally important to understand the nature of the dark sector. Finally, multiple observations also allow to test for consistency and systematic effects.

{As mentioned in Subsection \ref{subsec:euclid}, the impact of the addition of standard sirens from the Einstein Telescope to the LSS data of a \euclid-like survey was found to be very modest for the models here studied. This is, however, not necessarily true for more general models, as discussed above, and particularly for other planned interferometers. As shown in \cite{Baker2021}, LISA could improve the bound on the Planck mass running rate in Horndeski theories from CMB+BAO+RSD by a factor 5. Therefore, a natural next step would be to compare measurements of the GW friction from the network of current and future GW detectors with slip measurements from LSS, for more general viable models, for example relaxing the condition $\alpha_M = - \alpha_B$.} Furthermore, we stress that our results from the GW side could change a bit with the exact configuration of the interferometer or with careful selection of the GW events. Therefore, our conclusions follow for  tests of the distance-redshift relation with standard sirens performed as described in section \ref{sec:data}, but could be different for other kinds of analysis with standard sirens.

{In this work, we also analysed the constraints from LSS and GWs on the no-slip model, an interesting case of MG that produces modifications in the GW friction but does not generate gravitational slip, being a blind spot in tests of gravity through measurements of the latter. Assuming the Universe is well described by a no-slip model as far as possible
from $\Lambda$CDM but compatible with CMB data, we found that LSS also provides stronger bounds in this case.}

Here, the cosmological background was fixed to that of $\Lambda$CDM in all cases. A second analysis could be made, then, parametrizing the dark energy equation of state $w_{\textrm{DE}}$ in order to include background effects.  On the GW side, however, our simplification is reasonable since, despite the likelihood dependency on the EM luminosity distance, it has been shown \cite{Zhao2011,Belgacem2018,Matos2021}  that this test provides only a weak constraint on $w_{\textrm{DE}}$, as compared to the one on the friction $\Xi$.

We comment that Eq.\ \eqref{gw_distance} shows that, in the end, the effect of MG in $\Xi$ depends only on the values of the Planck mass at the emission and detection events. This fact is also valid in a more general context of an arbitrary background metric \cite{Dalang2020}, and thus it naturally raises the question on whether standard sirens would only be sensitive to the local value of the Planck mass \cite{Lagos2020, Baker2021}. When interpreting effective field models as Horndeski theories, the deviations from GR on the local values of $M_{\ast}^2$ can be suppressed by some of the known screening mechanisms, which was not considered here. This is not the case for Vainshtein screening \cite{Kimura2012}, in which case $\alpha_M$ is subject to constraints from local tests of gravity.

In fact, this is one caveat that arises once interpreting our model in the framework of Horndeski theory. The constraint from lunar laser ranging on the variation of the Planck mass is $|\alpha_M| < 0.02$ \cite{Williams2004, Tsujikawa2019}, and any cosmological probe at present provides weaker constraints than this one, making it harder to construct a model with $\alpha_M \neq 0$ on cosmological scales. Furthermore, the assumption $\alpha_B = -\alpha_M$ implies $G_{3X} = 0$ on cosmological scales, as discussed in appendix \ref{app:horndeski}. However, $G_{3X} = 0$ should not hold at smaller scales such as the solar system because the $X$ dependence of $G_3$ is critical for the Vainshtein screening to work. Therefore, one should keep in mind at which smaller scales our model is still valid.

Outside the scope of Horndeski theories, there are other gravity theories in which $\Xi$ is modified not simply because of the variation of the Planck mass, such as the non-local models \cite{Belgacem2018}, which by itself justifies testing gravity with the redshift-GW-distance relation. We stress that our goal here is to compare \textit{large-scale} constraints on effective field theories for cosmology being agnostic to what happens in the small scales. We point out that the simultaneous MG effect on the propagation of GWs and on the anisotropic stress is a general feature that goes beyond the exact models that we chose to study,  which are simple prototypes to compare the constraining power of the two observables.

Finally, it is possible to show that our base model with $\Omega_0 < 0$ has ghost instabilities. However, since Planck prefers negative values of $\Omega_0$ we did not impose priors to avoid ghosts. Again, our goal is not to investigate these particular models, that sometimes have physical problems, but rather to use them as an example for the combination of slip and GW distance measurements.

\acknowledgments

It is a pleasure to thank Luca Amendola, Charles Dalang, Pierre Fleury, Michele Mancarella,  Miguel Quartin,  Matteo Biagetti, and Ioav Waga and for useful discussions. We also thank the anonymous referee for his/her contributions to the final version of this paper.

I.S.M.~thanks the Department of Theoretical Physics of the University of Geneva and its members for hospitality during the PhD scholarship funded by both Brazilian agency CAPES (No 88887.569351/2020-00) and the department. I.S.M. also thanks Brazilian funding agency CNPq for PhD scholarship GD 140324/2018-6. E.B.~has received funding from the European Union’s Horizon 2020 research and innovation programme under the Marie Skłodowska-Curie grant agreement No 754496. M.K.\ acknowledges funding from the Swiss National Science Foundation.

\appendix

\section{Horndeski gravity} \label{app:horndeski}

The Horndeski theories of gravity are defined by the action
\begin{equation}
S = \int \sqrt{-g}d^4x\left[\sum_{i=2}^5 \mathcal{L}_i + \mathcal{L}_m \right]\,,
\end{equation}
with
\begin{align*}
\mathcal{L}_2 &= G_2(\phi, X)\,,  \\
\mathcal{L}_3 &= -G_3(\phi, X)\Box \phi\,, \\
\mathcal{L}_4 &= G_4(\phi, X)R + G_{4X}(\phi, X)\left[ (\Box \phi)^2 - \phi_{;\mu \nu} \phi^{;\mu \nu} \right]\,, \\
\mathcal{L}_5 &= G_5(\phi, X)G_{\mu \nu}\phi^{;\mu \nu} - \frac{1}{6}G_{5X}(\phi, X) \left[ (\Box \phi)^3 + 2{\phi_{;\mu}}^{\nu}{\phi_{;\nu}}^{\alpha}{\phi_{;\alpha}}^{\mu} - 3\phi_{;\mu \nu}\phi^{;\mu \nu}\Box \phi \right]\,,
\end{align*}
where $G_2, \dots, G_5$ are arbitrary functions of the scalar field $\phi$ and its kinetic energy $X :=  -\frac{1}{2}\phi_{;\mu}\phi^{;\mu}$ and $\mathcal{L}_m$ is the Lagrangian of the matter content.

The effective Planck mass in these theories is defined as
\begin{equation}
    M_{\ast}^2 := 2(G_4 -2XG_{4X} + XG_{5\phi} - \dot{\phi}HXG_{5X})\,.
\end{equation}

As demonstrated in \cite{Bellini2014}, the four time-dependent property functions are enough to describe the evolution of the linear scalar and tensor perturbations in a FLRW background. They can be expressed in terms of the functions defining the Lagrangian as
\begin{align}
    \alpha_M &:= \frac{d \log M_{\ast}^2}{d \log a}\\
    \alpha_T &:= \frac{2X}{M_{\ast}^2}\left[ 2G_{4X} - 2G_{5\phi} - (\Ddot{\phi} - \dot{\phi}H)G_{5X}\right]\\
    \alpha_K &:= \frac{2X}{HM_{\ast}^2} \bigg[ G_{2X} + 2XG_{2XX} - 2G_{3\phi} - 2XG_{3\phi X} \nonumber \\ &+ 6\dot{\phi}H(G_{3X} + XG_{3XX} - 3G_{4\phi X} - 2XG_{4\phi XX}) \nonumber \\ & + 6H^2(G_{4X} + 8XG_{4XX} + 4X^2G_{4XXX} - G_{5\phi} - 5XG_{5\phi X} - 2X^2G_{5\phi XX}) \nonumber \\ &+ 2\dot{\phi}H^3(3G_{5X} + 7XG_{5XX} + 2X^2G_{5XXX}) \bigg] \\
    \alpha_B &:= \frac{2\dot{\phi}}{HM_{\ast}^2}(XG_{3X} - G_{4\phi} - 2XG_{4\phi X}) + \frac{8X}{M_{\ast}^2}(G_{4X} + 2HG_{4XX} - G_{5\phi} - XG_{5\phi X}) \\ & + \frac{2\dot{\phi}XH}{M_{\ast}^2}(3G_{5X} + 2XG_{5XX})
\end{align}

Once given a Lagrangian-based model $G$'s, the $\alpha$ functions as well as the evolution of the background can be computed. However, the way back is not trivial, and in general is difficult to design functions $G$'s that lead to a given effective cosmology ($\alpha$'s and background evolution). For the base model studied here, however, we can asses some properties of a corresponding Horndeski theory. The requirement of a GW speed equals to that of light, $\alpha_T = 1$, is naturally achieved, without the need of any subtle cancellation, whenever $G_{4X} = G_{5\phi} = G_{5X} = 0$ \cite{Baker2017, Creminelli2017, Ezquiaga2017, Sakstein2017}. By also making the second assumption defining our base models, i.e. $\alpha_B = -\alpha_M$, we conclude that they constitute a general class of conformally coupled models with the additional condition $G_{3X} = 0$.

Instead of parametrizing directly the property functions, here we worked alternatively with the EFT base functions $(\Omega, \gamma_1, \gamma_2, \gamma_3)$, whose correspondence with the $\alpha$'s can be found in \cite{Bellini2018}. In particular, when $\alpha_T = 0$ and $\alpha_B = - \alpha_M$, it follows that:
\begin{equation}
    \Omega = -1 + M_{\ast}^2\,, \quad \gamma_2 = \gamma_3 = 0\,, \quad
    \gamma_1 = \frac{1}{4a^2H_0^2}\left[\alpha_K M_{\ast}^2 \mathcal{H}^2 - \mathcal{C}\right]\,,
\end{equation}
where
\begin{equation}
    \mathcal{C} = (\mathcal{H}^2 - \mathcal{H}')M_{\ast}^2 (2 + \alpha_M)  - \mathcal{H}M_{\ast}^2\left(\alpha_M' - \mathcal{H}\alpha_M + \mathcal{H}\alpha_M^2 \right) - \frac{a^2(\rho_m + p_m)}{M_{pl}^2}\,.
\end{equation}

The choice of $\alpha_K$ that makes $c_s^2 = 1$ is such that $\gamma_1 = 0$.

{The $\mu$ and $\Sigma$ functions in eqs. (\ref{lensing}) can then be expressed in terms of the $\alpha$'s and the background evolution (see appendix of \cite{Alonso:2016suf}). In our particular models, in the quasi-static approximation, they become
\begin{equation}
    \mu = \frac{\eta}{M_{\ast}^2}\,, \quad \Sigma = \frac{\eta + 1}{2M_{\ast}^2}\,,
\end{equation}
where, from eq. (\ref{slip}), the slip is given by
\begin{equation}
    \eta = 1 - \frac{\alpha_M}{2 +(1 + w_m)\tilde{\rho}_{\textrm{m}}/H^2 - 2\alpha_M'/\mathcal{H}\alpha_M}.
\end{equation}
}

\section{Notes on the background evolution}

In this pedagogical appendix we show and comment on the equations regulating
the expansion history of the universe following different conventions
common in the literature. As we shall see, the aim is to clarify some
key concepts that can lead to confusion particularly in general modified
gravity theories. As in the main text of this paper, our framework
is the effective description of a single scalar field \textit{à la}
Horndeski in the spatially flat cosmological context. One possible convention, the one we use in this paper, is to write down the equations governing the evolution of the background
as
\begin{align}
 & 3M_{\textrm{Pl}}^{2}\mathcal{H}^{2}=a^{2}\left(\rho_{\textrm{m}}+\mathcal{E}_{\textrm{S}}\right)\label{eq:friedmann_1_conv_1}\\
 & M_{\textrm{Pl}}^{2}\left(2\mathcal{H}^{\prime}+\mathcal{H}^{2}\right)=-a^{2}\left(p_{\textrm{m}}+\mathcal{P}_{\textrm{S}}\right)\label{eq:friedmann_2_conv_1}\\
 & \rho_{\textrm{m}}^{\prime}+3\mathcal{H}\left(\rho_{\textrm{m}}+p_{\textrm{m}}\right)=0\label{eq:conservation_mat_conv_1}\\
 & \mathcal{E}_{\textrm{S}}^{\prime}+3\mathcal{H}\left(\mathcal{E}_{\textrm{S}}+\mathcal{P}_{\textrm{S}}\right)=0\,,\label{eq:conservation_de_conv_1}
\end{align}
where $M_{\textrm{Pl}}^{2}$ is the bare Planck mass (constant in
time), $a$ is the scale factor, $\mathcal{H}\equiv\frac{da}{ad\tau}$
is the conformal Hubble parameter, $\rho_{\textrm{m}}$ and $p_{\textrm{m}}$
are the density and pressure of the matter field respectively, and
$\mathcal{E}_{\textrm{S}}$ and $\mathcal{P}_{\textrm{S}}$ are the
density and pressure of the Horndeski field. Eqs.~(\ref{eq:friedmann_1_conv_1})
and (\ref{eq:friedmann_2_conv_1}) are the Friedmann equations, Eq.~(\ref{eq:conservation_mat_conv_1})
follows from the conservation of the stress-energy tensor of matter
and Eq.~(\ref{eq:conservation_de_conv_1}) from the conservation
of the stress-energy tensor of the Horndeski field. Note that only
three of these equations are independent, while the fourth can be
derived from the others. Here, in order to fix the expansion history
of the universe we need to specify two functions of time: these can
be $\rho_{\textrm{m}}$ and $\mathcal{E}_{\textrm{S}}$, and we can
get $\mathcal{H}$ from Eq.~(\ref{eq:friedmann_1_conv_1}); then
$p_{\textrm{m}}$ and $\mathcal{P}_{\textrm{S}}$ are determined from Eqs.~(\ref{eq:conservation_mat_conv_1})
and (\ref{eq:conservation_de_conv_1}) respectively. $\rho_{\textrm{m}}$
depends on the matter components we want to include in our system
(CDM, baryons, radiation, \ldots ), while $\mathcal{E}_{\textrm{S}}$
can be parameterized at our convenience. In this paper we chose to
fix it to $\Lambda$CDM, i.e.~$\mathcal{E}_{\textrm{S}}=\frac{\Lambda}{3}$,
where $\Lambda$ is the cosmological constant.

A second approach is to keep an explicit dependence of the background
equations on an effective Planck mass $M_{*}^{2}\left(\tau\right)$
defined as a function of the scalar field or parameterized as a function
of time. To do so, we can redefine the Horndeski density and pressure
as
\begin{align}
 & \hat{\mathcal{E}}_{\textrm{S}}\equiv\mathcal{E}_{\textrm{S}}+3\left(M_{*}^{2}-M_{\textrm{Pl}}^{2}\right)\frac{\mathcal{H}^{2}}{a^{2}}\\
 & \hat{\mathcal{P}}_{\textrm{S}}\equiv\mathcal{P}_{\textrm{S}}-\frac{1}{a^{2}}\left(M_{\textrm{*}}^{2}-M_{\textrm{Pl}}^{2}\right)\left(2\mathcal{H}^{\prime}+2\mathcal{H}^{2}\right)\,.
\end{align}
The Friedmann and conservation equations become
\begin{align}
 & 3M_{*}^{2}\mathcal{H}^{2}=a^{2}\left(\rho_{\textrm{m}}+\hat{\mathcal{E}}_{\textrm{S}}\right)\label{eq:friedmann_1_conv_2}\\
 & M_{\textrm{*}}^{2}\left(2\mathcal{H}^{\prime}+\mathcal{H}^{2}\right)=-a^{2}\left(p_{\textrm{m}}+\hat{\mathcal{P}}_{\textrm{S}}\right)\label{eq:friedmann_2_conv_2}\\
 & \hat{\mathcal{E}}_{\textrm{S}}^{\prime}+3\mathcal{H}\left(\hat{\mathcal{E}}_{\textrm{S}}+\hat{\mathcal{P}}_{\textrm{S}}\right)=\left(\rho_{\textrm{m}}+\hat{\mathcal{E}}_{\textrm{S}}\right)\mathcal{H}\alpha_{\textrm{M}}\label{eq:conservation_mat_conv_2}\\
 & \rho_{\textrm{m}}^{\prime}+3\mathcal{H}\left(\rho_{\textrm{m}}+p_{\textrm{m}}\right)=0\,,\label{eq:conservation_de_conv_2}
\end{align}
where
\begin{align}
\alpha_{\textrm{M}} & \equiv\frac{d\ln M_{*}^{2}}{d\ln a}=\frac{\left(M_{*}^{2}\right)^{\prime}}{M_{*}^{2}\mathcal{H}}\,.
\end{align}
It is possible to notice that within this set of equations, matter
is still conserved, but the Horndeski field is not. In particular,
it seems that within this second approach we have to fix one extra
function of time, i.e.~$M_{*}^{2}\left(\tau\right)$ and its derivative
$\alpha_{\textrm{M}}$, w.r.t.~Eqs.~(\ref{eq:friedmann_1_conv_1})-(\ref{eq:conservation_de_conv_1}).
Clearly, this must be an artifact: a simple redefinition can not change
the physics of the problem. By comparing these two conventions, one
has to conclude that all the information carried by $\hat{\mathcal{E}}_{\textrm{S}}$
and $M_{*}^{2}$ can be compressed into a single function of time,
i.e.~$\mathcal{E}_{\textrm{S}}$. This does not mean that one approach
is better than the other, as the physics behind is the same. Rather,
it is a question of personal taste and needs. It is only important
to keep the notation consistent also when computing the equations
for the evolution of the perturbations, and to have clear in mind
what each choice implies.

A third approach is to redefine both matter and DE as
\begin{align}
 & \tilde{\rho}_{\textrm{m}}=\frac{\rho_{\textrm{m}}}{M_{*}^{2}}\\
 & \tilde{p}_{\textrm{m}}=\frac{p_{\textrm{m}}}{M_{*}^{2}}\\
 & \hat{\mathcal{E}}_{\textrm{S}}\equiv\frac{\mathcal{E}_{\textrm{S}}}{M_{*}^{2}}+3\left(1-\frac{M_{\textrm{Pl}}^{2}}{M_{*}^{2}}\right)\frac{\mathcal{H}^{2}}{a^{2}}\\
 & \hat{\mathcal{P}}_{\textrm{S}}\equiv\frac{\mathcal{P}_{\textrm{S}}}{M_{*}^{2}}-\frac{1}{a^{2}}\left(1-\frac{M_{\textrm{Pl}}^{2}}{M_{*}^{2}}\right)\left(2\mathcal{H}^{\prime}+\mathcal{H}^{2}\right)\,.
\end{align}
With these redefinitions Eqs.~(\ref{eq:friedmann_1_conv_1})-(\ref{eq:conservation_de_conv_1})
become
\begin{align}
 & 3\mathcal{H}^{2}=a^{2}\left(\tilde{\rho}_{\textrm{m}}+\tilde{\mathcal{E}}_{\textrm{S}}\right)\\
 & 2\mathcal{H}^{\prime}+2\mathcal{H}^{2}=-a^{2}\left(\tilde{p}_{\textrm{m}}+\tilde{\mathcal{P}}_{\textrm{S}}\right)\\
 & \tilde{\rho}_{\textrm{m}}^{\prime}+3\mathcal{H}\left(\tilde{\rho}_{\textrm{m}}+\tilde{p}_{\textrm{m}}\right)=-\mathcal{H}\alpha_{\textrm{M}}\tilde{\rho}_{\textrm{m}}\\
 & \tilde{\mathcal{E}}_{\textrm{S}}^{\prime}+3\mathcal{H}\left(\tilde{\mathcal{E}}_{\textrm{S}}+\tilde{\mathcal{P}}_{\textrm{S}}\right)=\mathcal{H}\alpha_{\textrm{M}}\tilde{\rho}_{\textrm{m}}\,.
\end{align}
Here both matter and DE are not conserved but their sum is. The interaction
term, proportional to $\alpha_{\textrm{M}}$, shows that the two fluids
exchange energy. As in Eqs.~(\ref{eq:friedmann_1_conv_2})-(\ref{eq:conservation_de_conv_2}),
we still have one more function to specify w.r.t.~Eqs.~(\ref{eq:friedmann_1_conv_1})-(\ref{eq:conservation_de_conv_1}),
but the physics has to be the same. As a side comment, this approach
is particularly useful to show the real freedom of these theories,
i.e.~a constant rescaling of the Planck mass can be consistently
reabsorbed into a convenient redefinition of the densities, but it
is is very unpractical when trying to implement these equations into
existing Einstein-Boltzmann solvers. Indeed, one would have to modify
all the matter part of the code, while with the other two approaches
one can leave it untouched.

We conclude trying to clarify a couple of concepts:
\begin{itemize}
\item If we talk about ``background evolution'', what do we really mean?
Background is just measuring the expansion history of the universe,
i.e.~$\mathcal{H}$, a single time dependent function. Following
this, it makes sense to follow an approach were there is only one
DE function modifying the expansion history to remove degeneracies
in the parameter space.
\item What does it mean ``$\Lambda$CDM background''? Clearly, if $\alpha_{\textrm{M}}\neq0$,
requiring any of $\mathcal{E}_{\textrm{S}}$, $\hat{\mathcal{E}}_{\textrm{S}}$
or $\tilde{\mathcal{E}}_{\textrm{S}}$ to be constant (to mimick the
energy density of a cosmological constant) does not imply that the
others are. But, since what we observe is $\mathcal{H}$, we should
require the Hubble parameter to be $\Lambda$CDM, i.e.
\begin{align}
 & 3M_{\textrm{Pl}}^{2}\left(\frac{\mathcal{H}}{a}\right)_{\Lambda\textrm{CDM}}^{2}=a^{-3}\rho_{0\textrm{m}}+\frac{\Lambda}{3}\,,\label{eq:friedmann_lcdm}
\end{align}
where $\rho_{0\textrm{m}}$ is the energy density of matter today.
Here we restricted ourselves to the case were matter is composed solely
by CDM, but the generalization to other species is straightforward.
Stressing again that all the approaches presented here are valid and
consistent if properly implemented, from eq.~(\ref{eq:friedmann_lcdm})
it seems correct to identify $\mathcal{E}_{\textrm{S}}=\frac{\Lambda}{3}$
as the $\Lambda$CDM background rather than setting constant any of
the others.
\item In the effective approach to MG cosmologies that we follow here, the background evolution and the property functions $\alpha_M, \alpha_B, \alpha_K, \alpha_T$ are independent by construction. These are the independent degrees of freedom that completely specify the cosmological observables up to the linear level, regardless of how easy it is to find the corresponding Lagrangian within Horndeski theory. In general (e.g.\ if $\alpha_M\neq 0$) the perturbations will evolve differently from the $\Lambda$CDM case even if the expansion rate $H$ is the one of the cosmological standard model.
\end{itemize}

\bibliographystyle{JHEP}
\bibliography{gw_slip}

\end{document}